\documentclass[aps,prd,reprint,groupedaddress, showpacs]{revtex4-1}

\usepackage{times}
\usepackage {euscript}
\usepackage{amsthm}
\usepackage{graphicx}
\usepackage{amsmath}
\usepackage{amssymb}
\usepackage{amsfonts}
\usepackage[english]{babel}
\usepackage{epstopdf}
\usepackage{multirow,makecell,array}
\usepackage{mathtools}
\usepackage{color}
\usepackage[shortlabels]{enumitem}
\usepackage{float}

\begin{document}


\title{Locally-constant field approximation in studies \\ of electron-positron pair production in strong external fields}


\author{I.~A.~Aleksandrov$^{1, 2}$}
\author{G.~Plunien$^{3}$} \author{V.~M.~Shabaev$^{1}$}
\affiliation{\vspace{0.1cm}$^1$~Department of Physics, St. Petersburg State University, Universitetskaya Naberezhnaya 7/9, Saint Petersburg 199034, Russia \\$^2$~NRC ``Kurchatov Institute'', Akademika Kurchatova Square 1, Moscow 123182, Russia \\ $^3$~Institut f\"ur Theoretische Physik, Technische Universit\"at Dresden, Mommsenstrasse 13, Dresden D-01062, Germany
}


\begin{abstract}
In the present investigation we revisit the widely-used locally-constant field approximation (LCFA) in the context of the pair-production phenomenon in strong electromagnetic backgrounds. By means of nonperturbative numerical calculations, we assess the validity of the LCFA considering several spatially homogeneous field configurations and a number of space-time-dependent scenarios. By studying the momentum spectra of particles produced, we identify the criteria for the applicability of the LCFA. It is demonstrated that the Keldysh parameter itself does not allow one to judge if the LCFA should perform accurately. In fact, the external field parameters must obey less trivial relations whose form depends on the field configuration. We reveal several generic properties of these relations which can also be applied to a broader class of other pair-production scenarios.

\end{abstract}

\maketitle


\section{Introduction}\label{sec:intro}

Quantum electrodynamics (QED) incorporating strong external backgrounds predicts a number of remarkable nonlinear phenomena such as light-by-light scattering, vacuum birefringence, quantum radiation reaction, and the vacuum production of electron-positron pairs (see, e.g., Ref.~\cite{dipiazza_rmp_2012} for review). The latter process~\cite{sauter_1931,euler_heisenberg,schwinger_1951} is the focus of the present study. It is well known that sufficiently strong external fields cannot be treated by perturbation theory which makes the corresponding regime particularly intriguing. The need for nonperturbative methods represents a serious challenge for theorists. Since the exact calculations in the case of external fields varying both in space and time seem extremely complicated, it is strongly desirable to approximate a realistic field configuration by a simpler background. The spatiotemporal dependence of the external field can partially be taken into account if such a simplification is made locally and the results are then summed (averaged) over the space-time. This approach is commonly referred to as the locally-constant field approximation (LCFA). Let $E$ and $\omega$ be the characteristic external field strength and its frequency. To be able to employ the LCFA, one usually requires the pair-formation length $l_\text{c} = mc^2/(|eE|)$ be much less than the laser radiation wavelength $\lambda$. The condition $l_\text{c} \ll \lambda$ is equivalent to $\xi \gg 1$ where $\xi$ is the adiabaticity parameter defined as $\xi = |eE|/(mc\omega)$ (it is the inverse of the Keldysh parameter~\cite{keldysh}). Although this corresponds to the nonperturbative (Schwinger) regime, which is of major interest, it is still unclear to which extent one can rely on the LCFA results and whether $\xi \gg 1$ can be considered as a sufficient requirement. On the other hand, a very important role of the spatial inhomogeneities was recently reported in a number of studies regarding the pair-production phenomenon (see Refs.~\cite{aleksandrov_prd_2016, kohlfuerst_plb_2016, aleksandrov_prd_2017_2, aleksandrov_prd_2018, kohlfuerst_prd_2018, lv_pra_2018, torgrimsson_2018, peng_arxiv_2018, kohlfuerst_epjp_2018, karbstein_prd_2017}). In the present investigation, we examine the validity of the LCFA in order to find out which values of the external field parameters make the LCFA applicable to the corresponding problems.

We also note that the LCFA is frequently invoked for studying other strong-QED processes. In the past few years the validity of the LCFA was addressed in a number of investigations. For instance, in Ref.~\cite{meuren_prd_2016} the LCFA was elaborated in the context of the nonlinear Breit-Wheeler process. In Refs.~\cite{dipiazza_2018, blackburn_2018, ilderton_arxiv_2018} it was demonstrated that the LCFA may fail to properly predict the low-energy part of the photon spectrum in studies of nonlinear Compton scattering. This provides even further motivation for our present study.

We focus on the evaluation of the number density of particles produced and consider several space-time-dependent field configurations as well as several uniform backgrounds depending solely on time. The results obtained within the LCFA are compared to the exact spectra, i.e. momentum distributions calculated by taking into account the spatiotemporal dependence of the external field without any approximations. The nonuniform scenarios are examined by means of the nonperturbative numerical technique described in Ref.~\cite{aleksandrov_prd_2016}. Benchmarking the LCFA results against the corresponding precise values, we analyze the validity of this approximation.

The paper is organized as follows. In Sec.~\ref{sec:uniform} three different time-dependent field configuration are considered. In Sec.~\ref{sec:space-time} we turn to the analysis of several spatially inhomogeneous backgrounds. Finally, in Sec.~\ref{sec:discussion} we provide a discussion. Relativistic units ($\hbar = 1$, $c = 1$) are used throughout the paper.

\section{Spatially uniform fields}\label{sec:uniform}

In this section we discuss how one can employ the LCFA in the case of a purely time-dependent background. We assume that the external electric field of linear polarization vanishes outside the interval $[t_\text{in}, t_\text{out}]$. The main idea is to split this range into $N$ subintervals and approximate the field by a piecewise constant function: $E(t) = E_i$ for $t \in [t_i, t_{i+1}]$. After that one can sum all of the individual contributions arising from each subinterval. This approach will be attested by comparing its predictions to the exact values of the pair-production probabilities which can be extracted from two special sets of the {\it in} and {\it out} one-particle solutions of the Dirac equation. These solutions are determined by their asymptotic behavior at $t=t_\text{in}$ and $t=t_\text{out}$, respectively. Propagating a given {\it out} solution backwards in time and projecting it onto the {\it in} basis, one evaluates the number density of particles corresponding to this particular final state. This approach is described numerous times in literature (see, e.g., Ref.~\cite{fradkin_gitman_shvartsman}) and implemented in our study.

Since the LCFA approximates the external field within each subinterval by a constant profile, it is essential to examine first a simple case of a rectangular-like background. To begin with, we perform the exact calculations and identify the qualitative and quantitative patterns of the momentum distributions of particles created.

\subsection{Rectangular profile}\label{sec:uniform_rect}

The external field is assumed to have the form $E_x(t) = E_0 \theta(T/2 - |t|)$, $E_y=E_z=0$ ($t_\text{out}=-t_\text{in}=T/2$), where the parameters $E_0 > 0$ and $T$ are to be varied. The spectrum of particles produced depends only on longitudinal momentum projection $p_\parallel=p_x$ and transversal projection $p_\perp = \sqrt{p_y^2 + p_z^2}$. A nonzero transversal momentum effectively changes the electron mass, so that the pair $(m, p_\perp)$ is equivalent to $(\pi_\perp, 0)$ where $\pi_\perp = \sqrt{m^2 + p_\perp^2}$. The spectrum also does not depend on spin quantum number $s$. The number density of particles created per unit volume will be denoted by $n_{\boldsymbol{p},s}$, i.e.
\begin{equation}
n_{\boldsymbol{p},s} = \frac{(2\pi)^3}{V} \, \frac{d N_{\boldsymbol{p},s}}{d^3 \boldsymbol{p}}.\label{eq:density_notation}
\end{equation}
It turns out that the pair-production probabilities can be found exactly and expressed in terms of the Weber parabolic cylinder functions~\cite{gav_git_prd_1996} (see also Refs.~\cite{nikishov_jetp_1970, bagrov_jetp_1975, adorno_2018}). The corresponding exact relations yield exactly the same results as our numerical procedures.

\begin{figure*}[t]
\center{\includegraphics[height=0.23\linewidth]{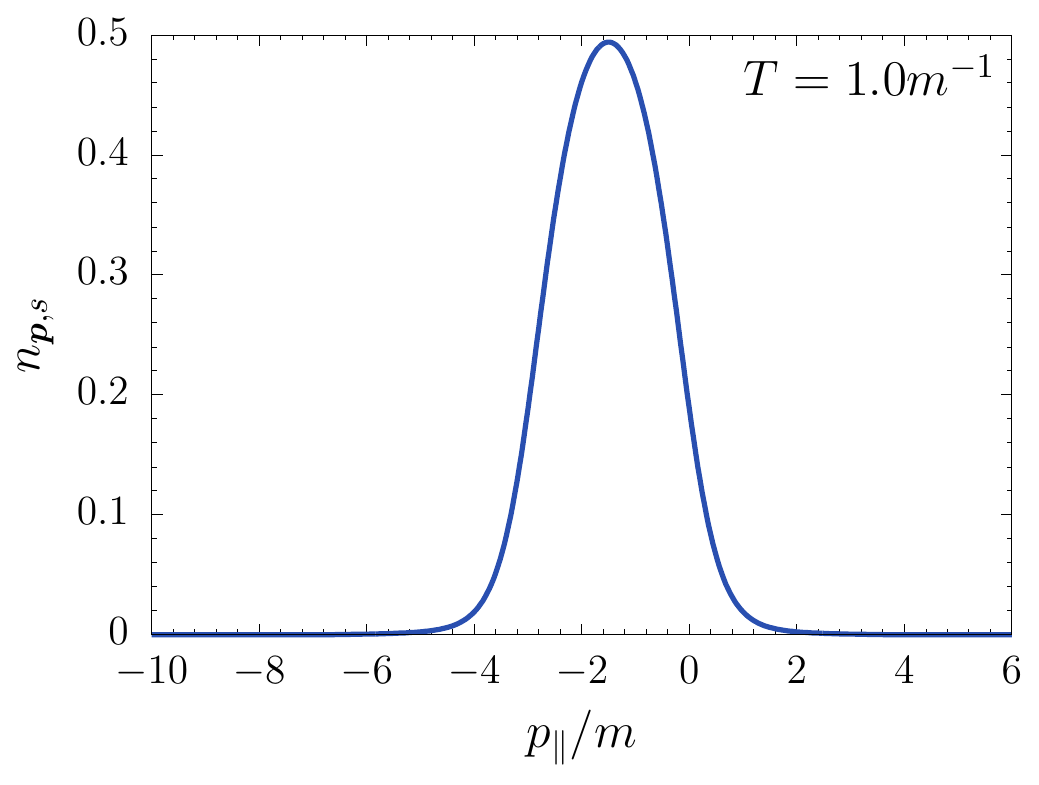}~~~\includegraphics[height=0.23\linewidth]{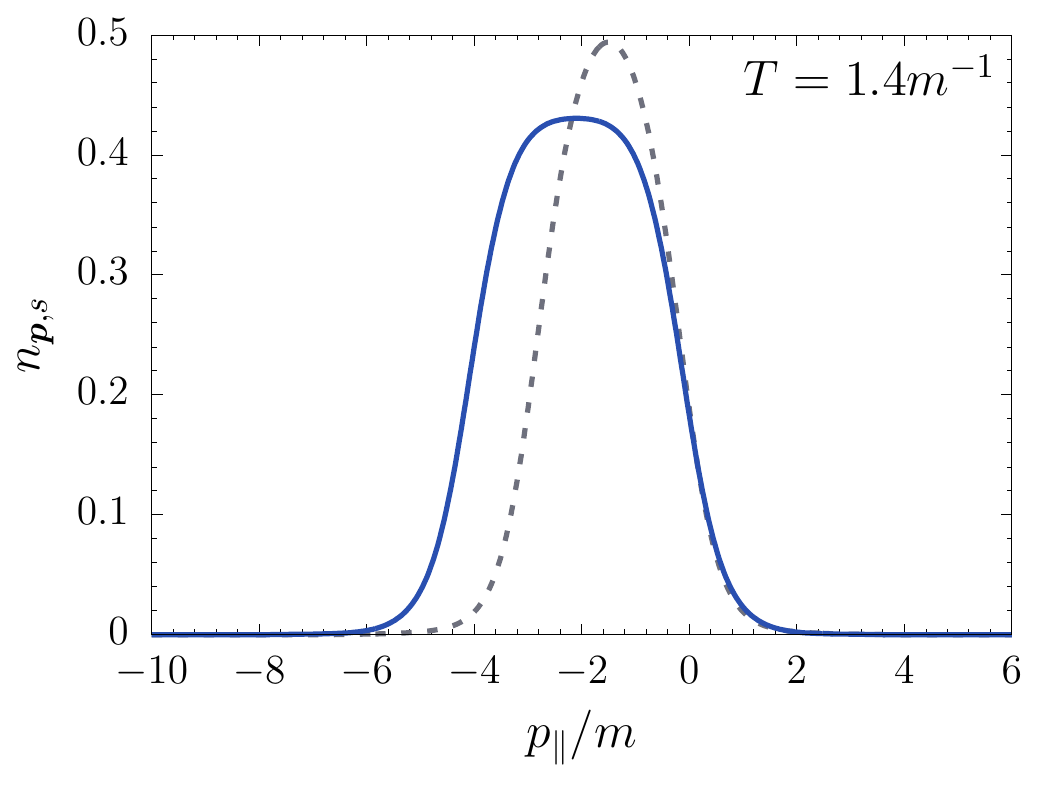}~~~\includegraphics[height=0.23\linewidth]{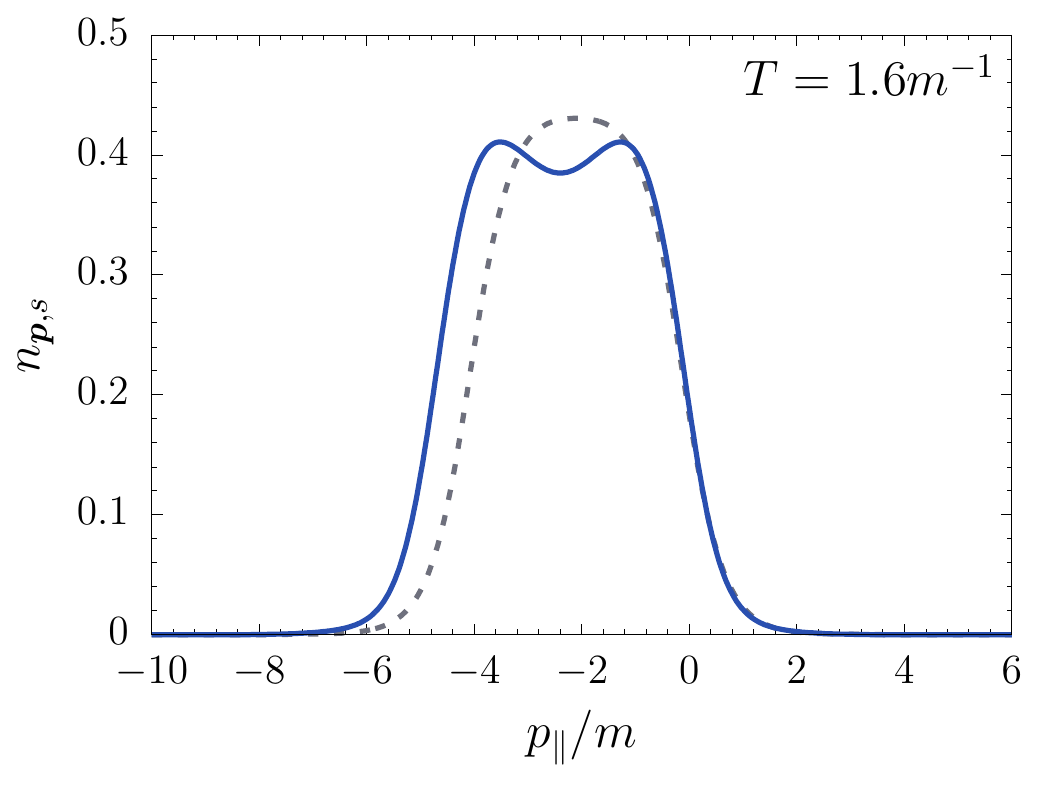}}\\
\center{\includegraphics[height=0.23\linewidth]{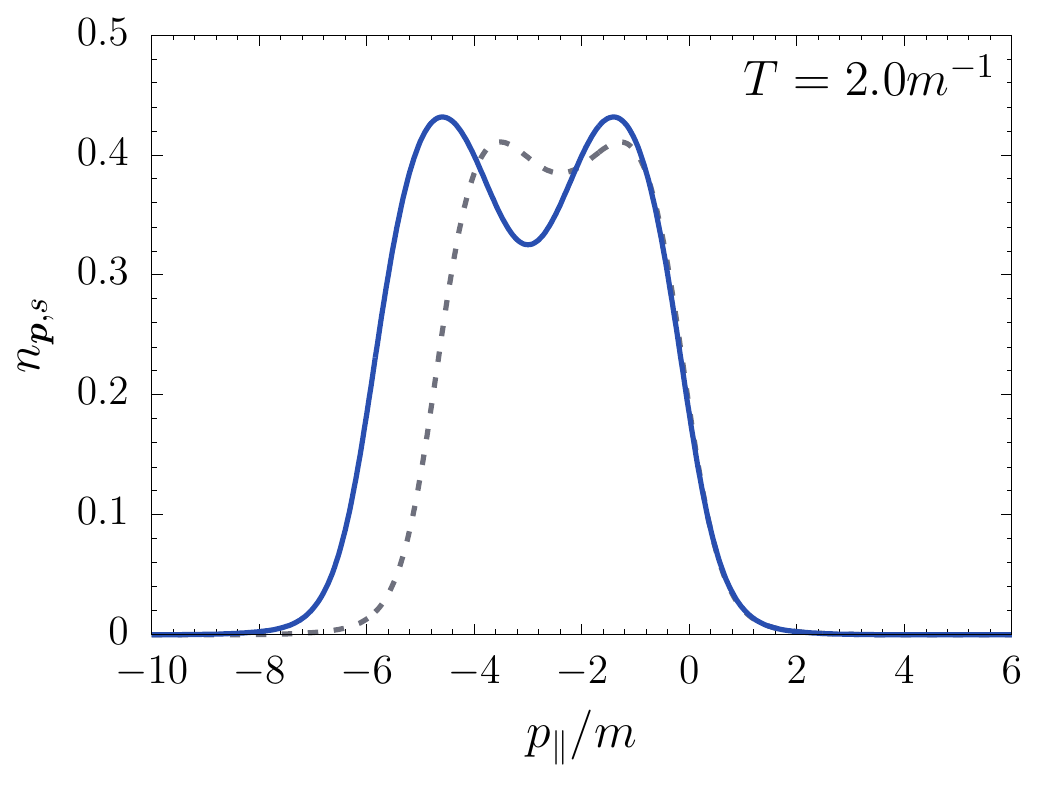}~~~\includegraphics[height=0.23\linewidth]{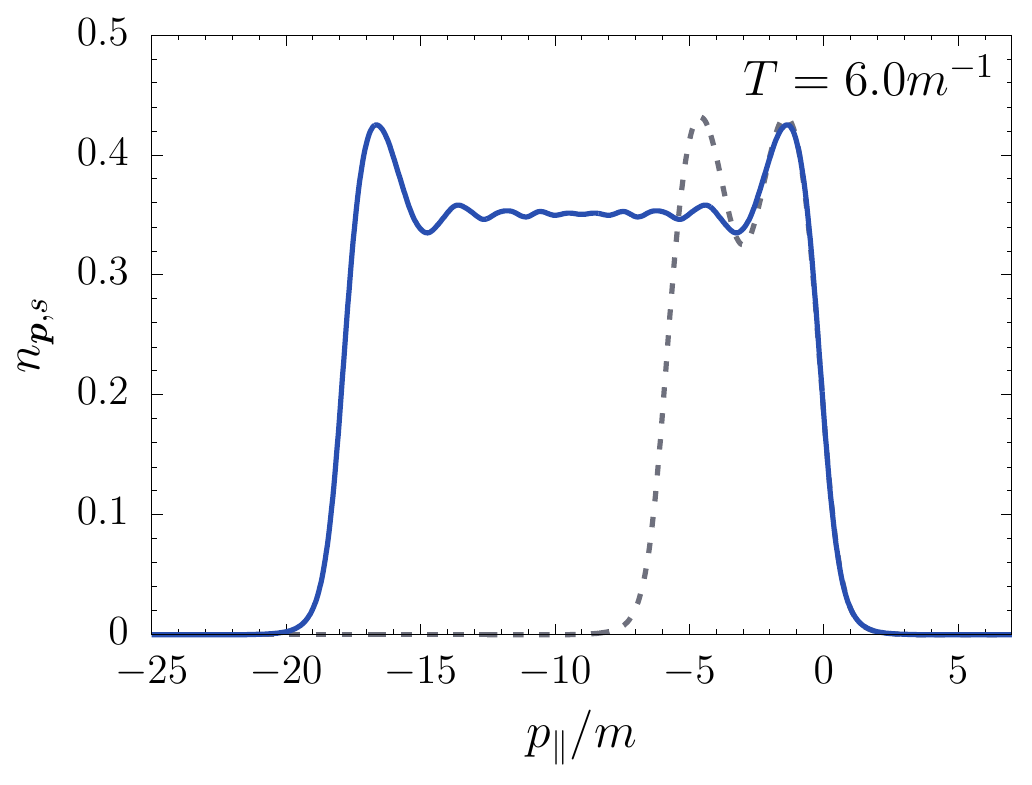}~~~\includegraphics[height=0.23\linewidth]{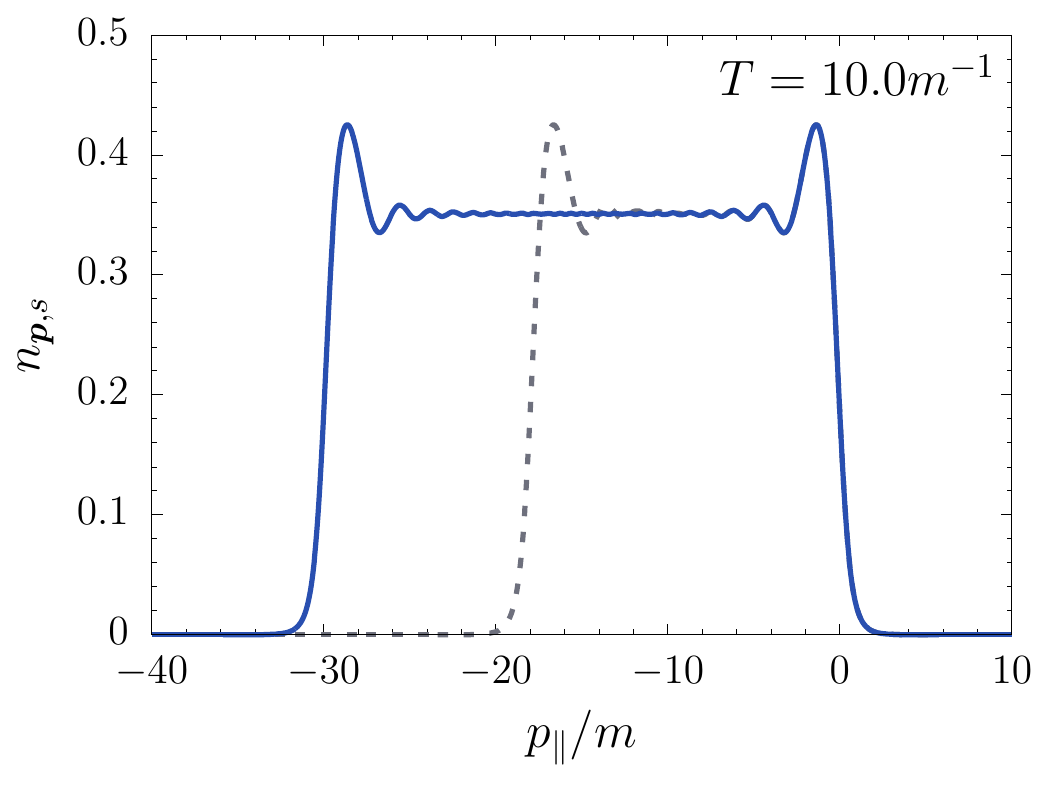}}
\caption{The momentum spectra of electrons created with $p_\perp = 0$ in the case of a rectangular-like electric field with $E_0 = 3E_\text{c}$ and various values of the pulse duration $T$. The dashed curve represents the spectrum for the previous value of $T$.}
\label{fig:rect_ex}
\end{figure*}

In order to make the following discussion clearer, we begin with an example of the $p_\parallel$ distribution of electrons for $E_0=3E_\text{c}$, $p_\perp = 0$, and various values of $T$ (see Fig.~\ref{fig:rect_ex}).
One observes a number of distinctive features. First, the momentum distribution takes a rectangular-like shape for sufficiently large $T$ and its width approximately equals $|e|E_0T$. Note that the results are expressed in terms of the kinetic momentum. Since the electron produced is being then accelerated by the external field opposite to the $x$ axis, the spectrum mostly lies in the negative-$p_\parallel$ region. Second, the momentum distribution gains a plateau region whose height corresponds to the Schwinger value
\begin{equation}
n_{\boldsymbol{p},s}^{\text{(Schwinger)}} = \mathrm{e}^{-\pi \lambda_{\boldsymbol{p}}(E_0)},~~\text{where}~~\lambda_{\boldsymbol{p}}(E) = \frac{\pi_\perp^2}{|e|E}.\label{eq:T_schwinger}
\end{equation}
In this particular case, it amounts to $0.351$. Third, the large-$T$ curves possess wiggles at the edges which represent the effects of the finite duration of the external electric pulse. These wiggles should be analyzed in more detail as the particles are likely to be produced with low kinetic energy and the main contribution from each interval $[t_i, t_{i+1}]$ will accordingly arise from the small-$p_\parallel$ parts of the spectra.

\begin{figure*}[t]
\center{\includegraphics[height=0.23\linewidth]{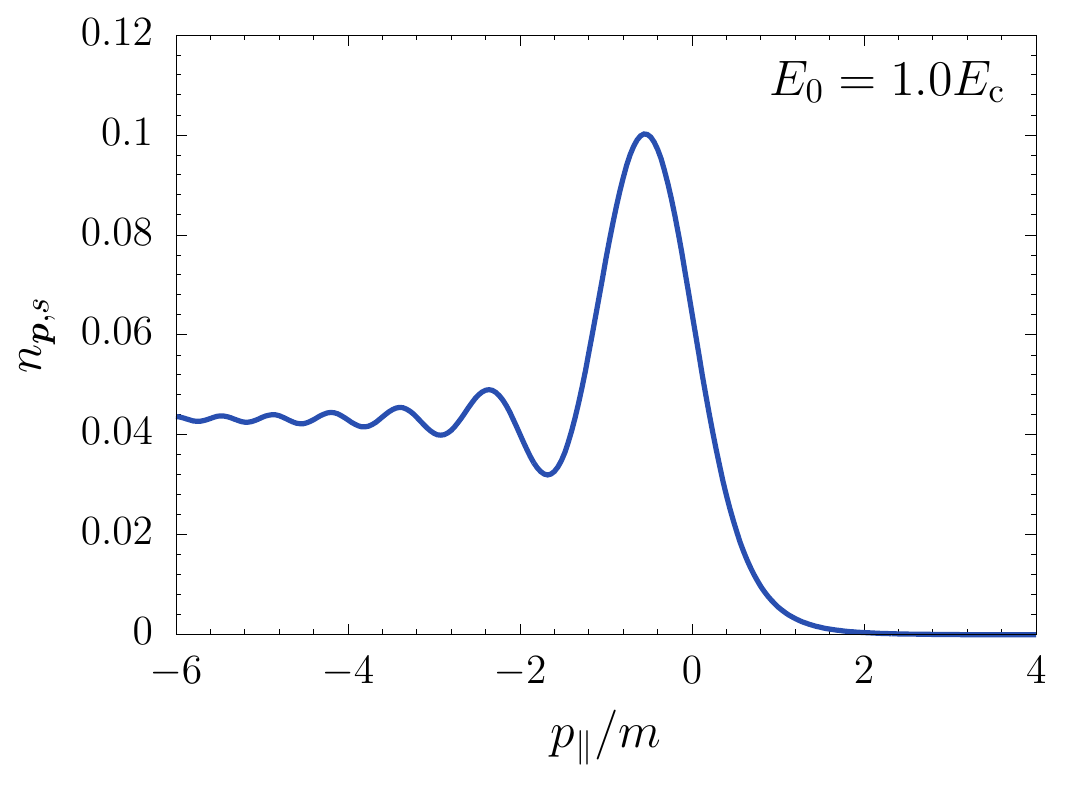}~~~\includegraphics[height=0.23\linewidth]{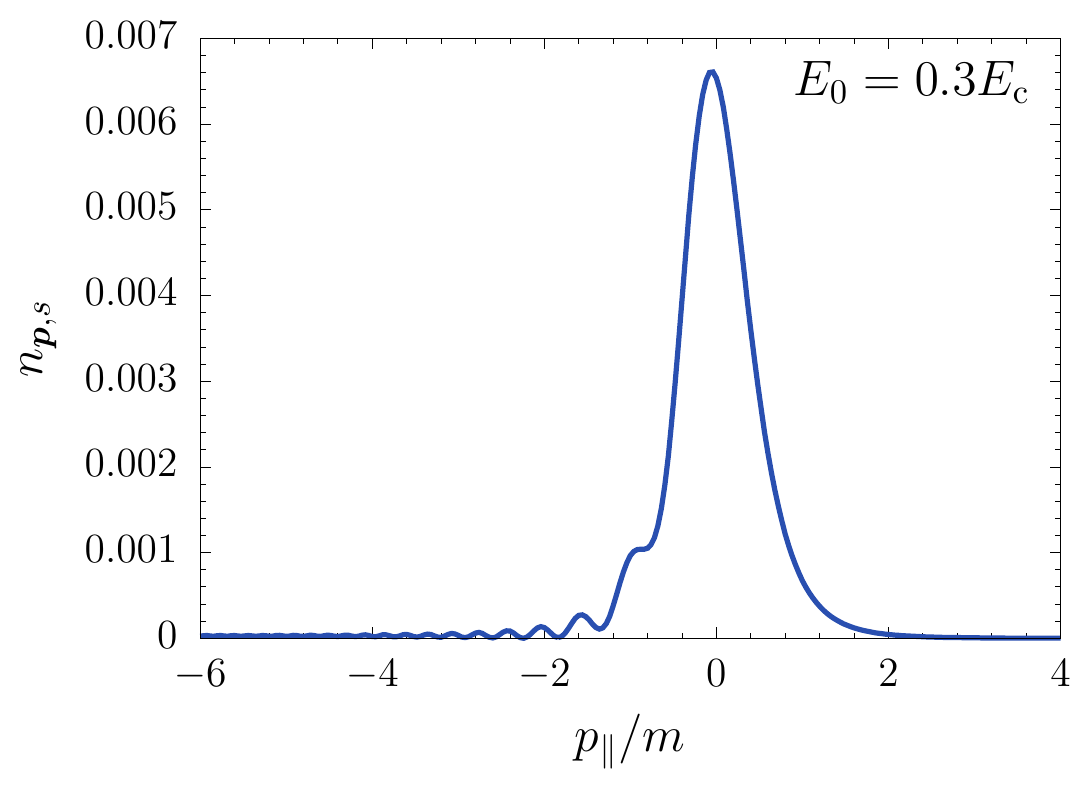}~~~\includegraphics[height=0.23\linewidth]{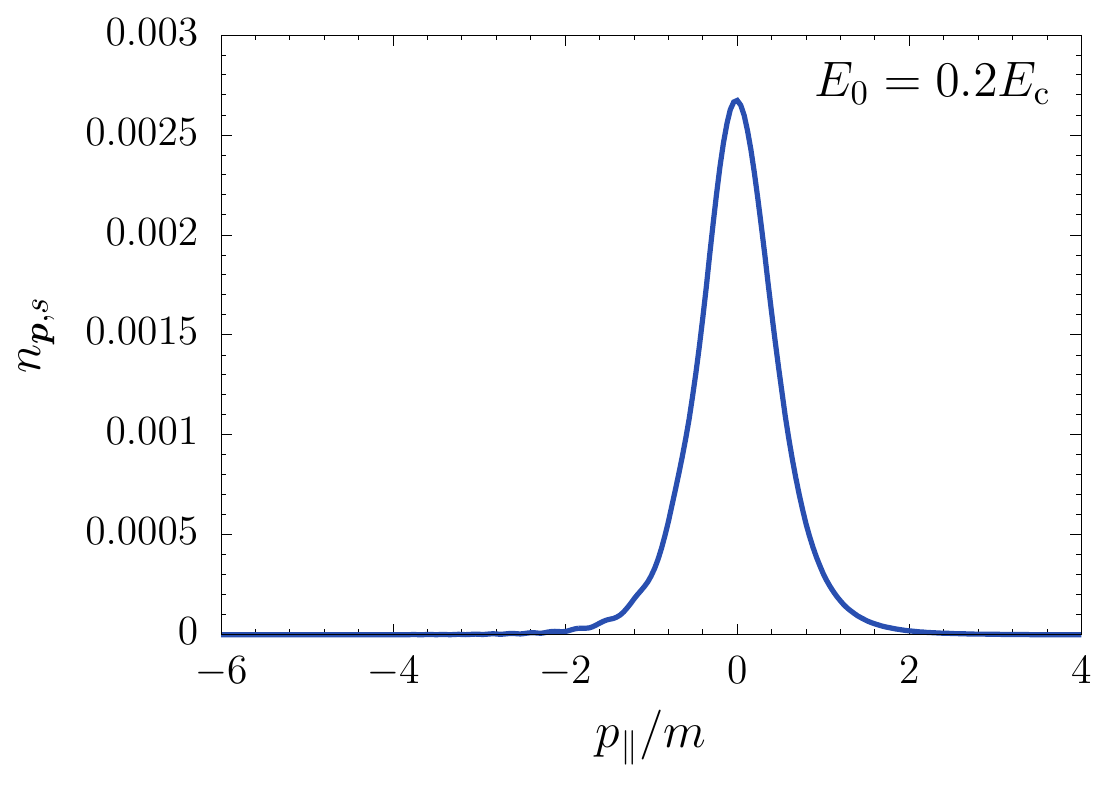}}
\caption{The momentum distributions in the case of a rectangular-like electric field with various values of $E_0$ ($p_\perp = 0$). The pulse duration is sufficiently large, so this part of the spectrum no longer depends on $T$.}
\label{fig:rect_smaller_E0}
\end{figure*}

We now present a quantitative description of the momentum distribution in the vicinity of $p_\parallel = 0$. We choose a sufficiently large value of $T$, so that the wiggles are already frozen, and perform the calculations for smaller values of $E_0$ (see Fig.~\ref{fig:rect_smaller_E0}).
One discovers that the spectrum becomes essentially an even function of $p_\parallel$ having a maximum at $p_\parallel = 0$ and negligible value of the Schwinger plateau. The graphs demonstrate that for small $E_0$ the pair-production process is entirely governed by the switching-on and -off effects. To further elaborate this issue, we present the ratio $\kappa = n_{\boldsymbol{p},s}/n_{\boldsymbol{p},s}^{\text{(Schwinger)}}$ at $p_\parallel = 0$ as a function of $E_0$ (see Fig.~\ref{fig:ratio}).
The pulse duration chosen is always sufficiently large so that the ratio is converged. It is seen that the finite-duration effects predominate over the infinite-pulse results once $E_0 \lesssim E_\text{c}$. Some other aspects concerning the switching-on and -off effects in the case of a rectangular-like pulse can be found in Ref.~\cite{adorno_2018}.

\subsection{LCFA for uniform fields}\label{sec:LCFA_uniform_procedure}

\begin{figure}[H]
\center{\includegraphics[width=0.95\linewidth]{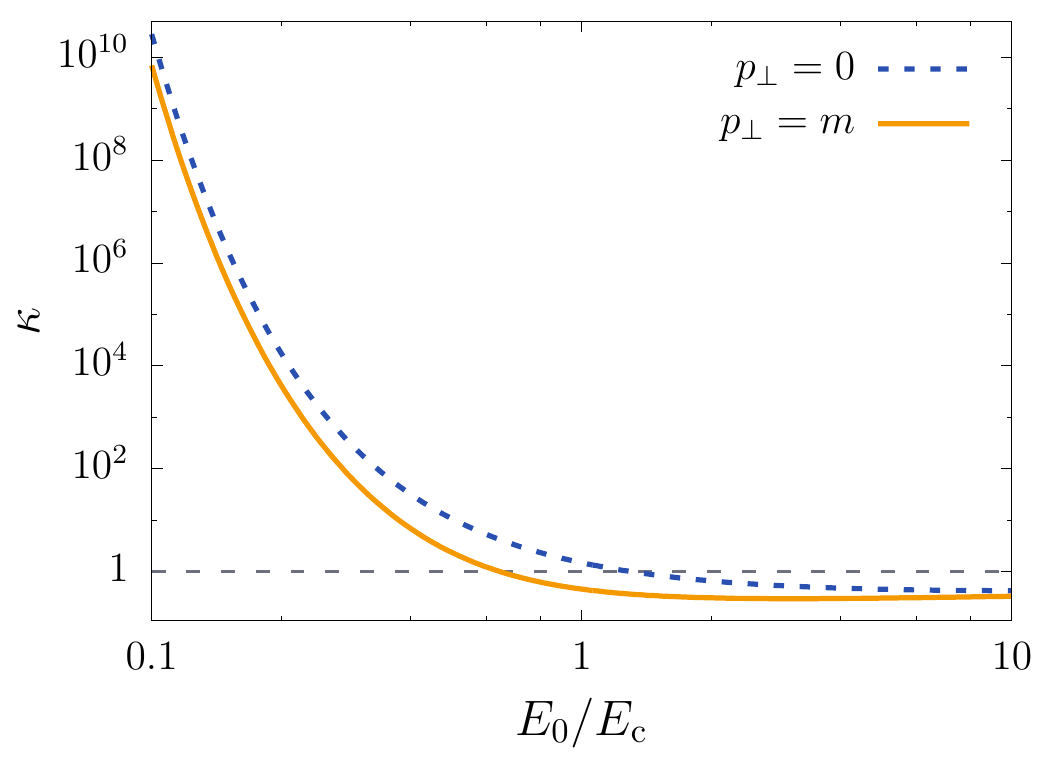}}
\caption{The ratio $\kappa = n_{\boldsymbol{p},s}/n_{\boldsymbol{p},s}^{\text{(Schwinger)}}$ at $p_\parallel = 0$ as a function of $E_0$ for two different values of $p_\perp$. The external field has a rectangular profile.}
\label{fig:ratio}
\end{figure}

Let us now discuss how one can employ the LCFA (for calculating the total amount of particles, this procedure is described, e.g., in Ref.~\cite{gavrilov_prd_2017}). For a general time-dependent background, we divide the time interval $[t_\text{in}, t_\text{out}]$ into $N$ subintervals: $t_k = t_{k-1} + \Delta t_k$, $k = 1,..., N$, $t_0=t_\text{in}$, $t_N=t_\text{out}$. In order to evaluate the mean number of particles produced with (final) kinetic momentum $\boldsymbol{p}$, we propagate it backwards in time according to $(\boldsymbol{p}_k)_\parallel = p_\parallel - e[A(t_k) - A(t_\text{out})]$, $(\boldsymbol{p}_k)_\perp = p_\perp$ and sum the individual contributions $n_{\boldsymbol{p}_k,s}$. One should then decide how to evaluate $n_{\boldsymbol{p}_k,s}$. It is now clear that the predominance of the finite-duration effects revealed in Figs.~\ref{fig:rect_smaller_E0} and \ref{fig:ratio} does not allow one to use the exact value for a static electric background of finite duration from Ref.~\cite{gav_git_prd_1996}. Accordingly, setting $n_{\boldsymbol{p}_k,s} = \mathrm{e}^{-\pi \lambda_{\boldsymbol{p}_k}[E(t_k)]}$ for $eE(t_k) \Delta t_k \leq (\boldsymbol{p}_k)_\parallel \leq 0$ in the limit $\Delta t_k = \Delta t \to 0$, one obtains the following expression for the total value of the number density in the case of a rectangular field profile:
\begin{equation}
n_{\boldsymbol{p},s}^{\text{(LCFA)}} =
\begin{cases}
\mathrm{e}^{-\pi \lambda_{\boldsymbol{p}}[E(t_*)]} &\text{if}~~p_\parallel \in [eE_0 T, 0],\\
0 &\text{otherwise},
\end{cases} \label{eq:lcfa_rect}
\end{equation}
where $t_*$ is the time instant when the longitudinal kinetic momentum vanishes: $p_\parallel (t_*) = p_\parallel - e[A(t_*) - A(t_\text{out})] = 0$. It yields
\begin{equation}
t_* = \frac{T}{2} - \frac{p_\parallel}{eE_0},\quad E(t_*) = E_0. \label{eq:lcfa_rect_t_star}
\end{equation}
Since $t_* \in [-T/2, T/2]$, the projection $p_\parallel$ should obey $eE_0 T < p_\parallel < 0$ as shown in Eq.~(\ref{eq:lcfa_rect}). This approach approximates the momentum spectrum by a rectangular of height $\mathrm{e}^{-\pi \lambda_{\boldsymbol{p}}(E_0)}$ and width $|e|E_0 T$. Although it does not reproduce the effects of the temporal finiteness of the external pulse, one can expect the LCFA to perform well in the case of more realistic configurations being switched on and off smoothly. Next we will consider the Sauter temporal dependence.

\subsection{Sauter pulse}\label{sec:uniform_sauter}

The external field has now the form
\begin{equation}
E_x (t) = \frac{E_0}{\cosh^2 (t/\tau)},\quad E_y = E_z = 0,\label{eq:sauter_field}
\end{equation}
where $\tau$ governs the pulse duration while $t_\text{in/out} \to \mp \infty$. The LCFA predicts the following value of the number density:

\begin{equation}
n_{\boldsymbol{p},s}^{\text{(LCFA)}} =
\begin{cases}
\mathrm{e}^{-\pi \lambda_{\boldsymbol{p}}[E(t_*)]} &\text{if}~~p_\parallel \in [2eE_0 \tau, 0],\\
0 &\text{otherwise},
\end{cases} \label{eq:lcfa_sauter}
\end{equation}
where $t_*$ obeys
\begin{equation}
\tanh \frac{t_*}{\tau} = 1 - \frac{p_\parallel}{eE_0\tau}. \label{eq:lcfa_sauter_t_star}
\end{equation}
Hence, within the region $p_\parallel \in (2eE_0 \tau, 0)$,
\begin{equation}
n_{\boldsymbol{p},s}^{\text{(LCFA)}} = \mathrm{exp}\bigg [ - \frac{\pi \pi_\perp^2 e E_0 \tau^2}{p_\parallel (p_\parallel - 2eE_0 \tau)}\bigg ].\label{eq:lcfa_sauter_explicit}
\end{equation}
This expression is to be compared with the exact result~\cite{narozhny_1970, gav_git_prd_1996}
\begin{widetext}
\begin{equation}
n_{\boldsymbol{p},s}^{\text{(exact)}} = \frac{\sinh \Big [ \frac{1}{2} \pi \tau (2e E_0 \tau + \omega_- - \omega_+)\Big ] \sinh \Big [ \frac{1}{2} \pi \tau (2e E_0 \tau + \omega_+ - \omega_-)\Big ]}{\sinh(\pi \omega_+ \tau) \sinh(\pi \omega_- \tau)},\label{eq:lcfa_sauter_exact}
\end{equation}
\end{widetext}
where $\omega_{\pm} = \sqrt{\pi_\perp^2 + (P_\parallel \mp eE_0\tau)^2}$ and $P_\parallel = p_\parallel - eE_0 \tau$.

\begin{figure}[b]
\center{\includegraphics[width=0.95\linewidth]{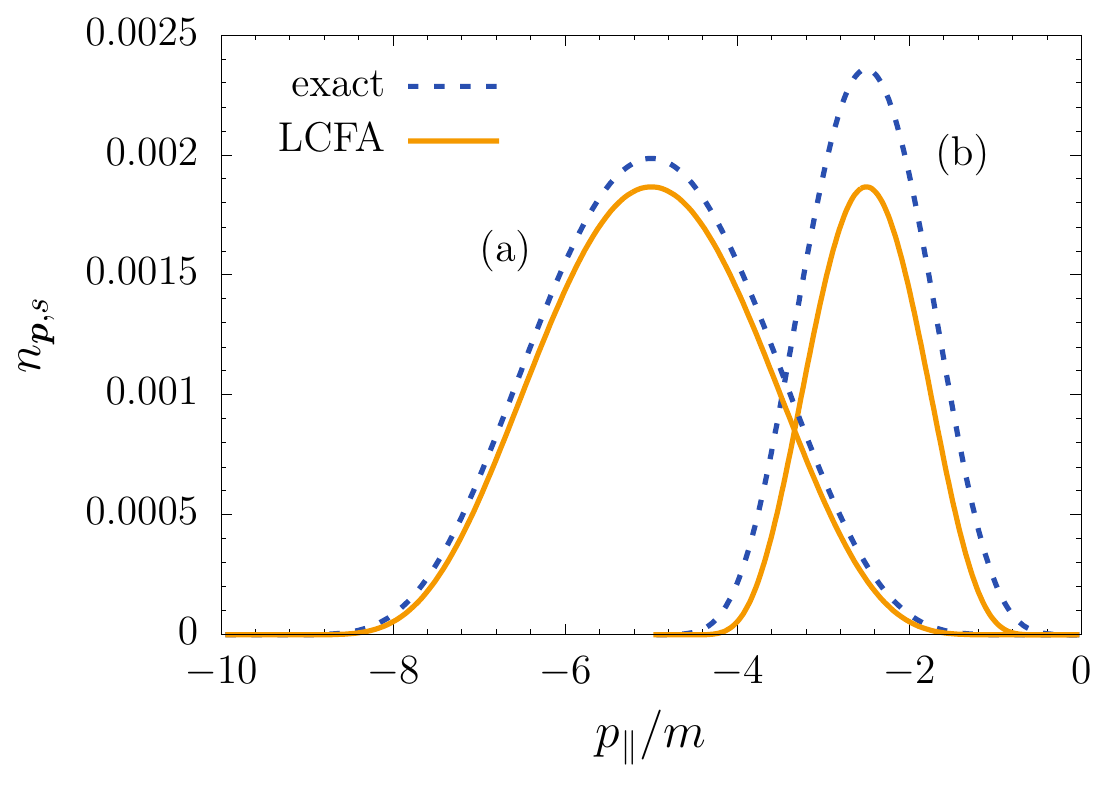}}
\caption{The momentum spectra of particles created by the Sauter pulse~(\ref{eq:sauter_field}) with $E_0 = 0.5 E_\text{c}$ ($p_\perp = 0$). The dashed lines represent the exact results while the solid lines correspond to the LCFA estimates. The pulse duration is (a)~$\tau = 10 m^{-1}$ and (b)~$\tau = 5 m^{-1}$.}
\label{fig:sauter_spectra}
\end{figure}

\begin{figure}[b]
\center{\includegraphics[width=0.95\linewidth]{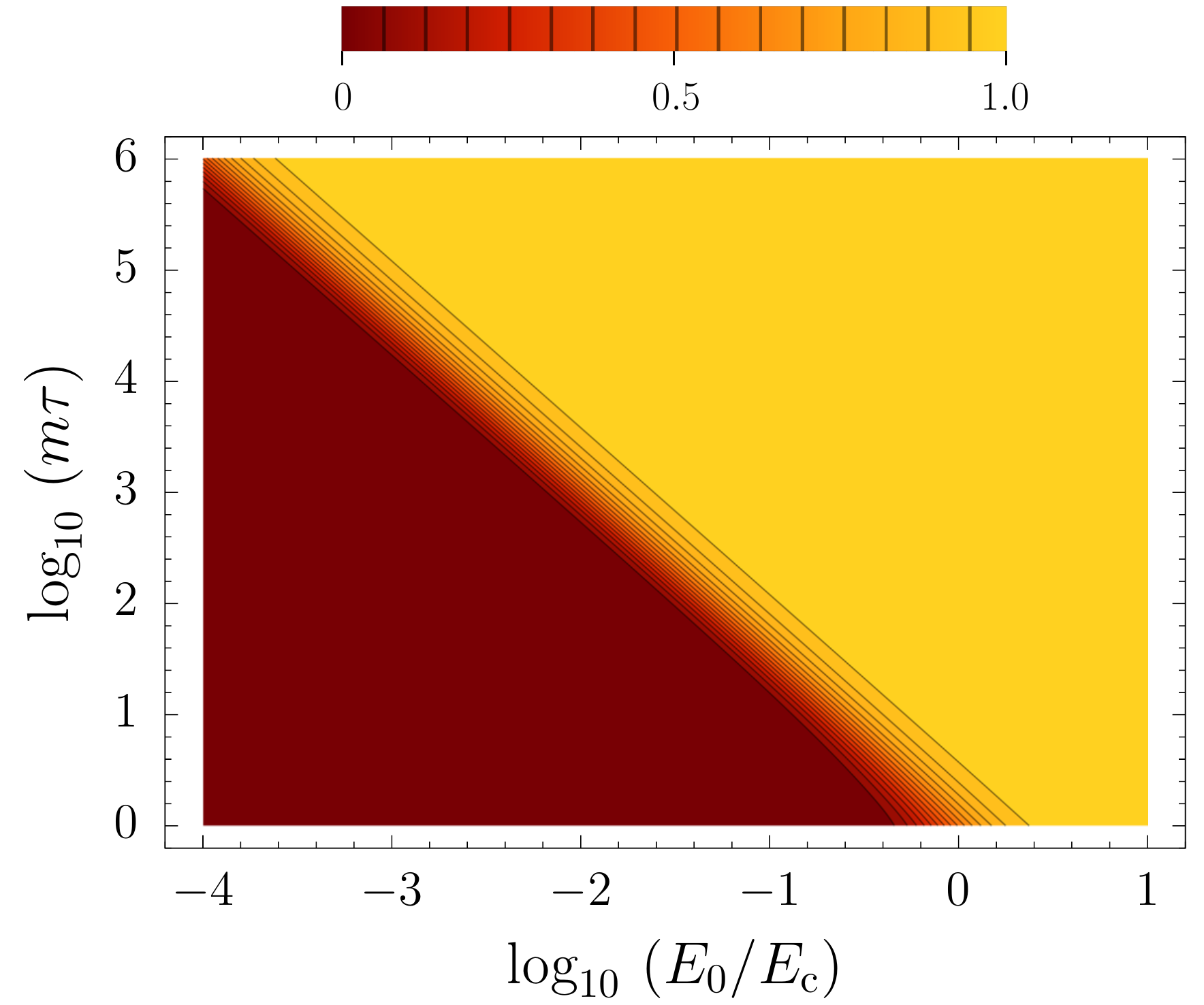}}
\caption{The ratio $\zeta = n_{\boldsymbol{p},s}^{\text{(LCFA)}}/n_{\boldsymbol{p},s}^{\text{(exact)}}$ at $p_\parallel = eE_0 \tau$ ($P_\parallel = 0$) and $p_\perp = 0$ as a function of $E_0$ and $\tau$ in the case of the Sauter field configuration~(\ref{eq:sauter_field}).}
\label{fig:sauter_zeta}
\end{figure}

In Fig.~\ref{fig:sauter_spectra} we present the momentum spectra computed by means of Eqs.~(\ref{eq:lcfa_sauter_explicit}) and (\ref{eq:lcfa_sauter_exact}), respectively, for $p_\perp = 0$, $E_0 = 0.5 E_\text{c}$, and two different values of $\tau$.
Our analysis indicated that for sufficiently small $\tau$, the LCFA can substantially underestimate the pair-production probabilities. Let us consider the ratio $\zeta = n_{\boldsymbol{p},s}^{\text{(LCFA)}}/n_{\boldsymbol{p},s}^{\text{(exact)}}$ at $p_\parallel = eE_0 \tau$ ($P_\parallel = 0$) and $p_\perp = 0$ as a measure of this underestimation (this value of the momentum projection corresponds to the maximal number density). We also make a realistic assumption $m \tau \gg 1$. It follows that for $\xi \equiv |e|E_0\tau/m \gg 1$,
\begin{equation}
\zeta = \mathrm{exp} \bigg ( \! -\frac{\pi m\tau}{4\xi^3} \Big [ 1 + \mathcal{O} (1/\xi^2) \Big ] \bigg ).\label{eq:lcfa_sauter_zeta}
\end{equation}
Therefore, one should mind that
\begin{equation}
\frac{m\tau}{\xi^3} \ll 1~~\Longleftrightarrow~~|eE_0|^{3/2} \tau \gg m^2.\label{eq:lcfa_sauter_condition}
\end{equation}
The condition derived is stronger than mere $\xi \gg 1$, so the criterion of the LCFA justification turns out to be quite nontrivial. In Fig.~\ref{fig:sauter_zeta} we display the ratio $\zeta$ as a function of $E_0$ and $\tau$.
The border between the regions with $\zeta = 0$ and $\zeta = 1$ clearly confirms the condition (\ref{eq:lcfa_sauter_condition}) (e.g., the line $\zeta = 0.9$ corresponds to $|eE_0|^{3/2} \tau \approx 2.6 m^2$).

In addition, we point out that in the range $E_0 \ll E_\text{c}$, one can also employ the imaginary time method (ITM)~\cite{brezin_1970, popov_1972, marinov_1972}. Unlike the LCFA, which directly sums the particle yields arising from each time interval, the ITM is based on the calculation of the imaginary part of the classical action along the tunneling trajectory. The ITM accurately reproduces the exact result (\ref{eq:lcfa_sauter_exact}) provided $m\tau \xi \gg 1$~\cite{popov_1972}. This means that in the case of small field amplitudes, the ITM has a broader applicability than that of the LCFA [one needs to satisfy $|eE_0|^{1/2} \tau \gg 1$ instead of (\ref{eq:lcfa_sauter_condition})]. However, as $E_0$ approaches the Schwinger limit, the LCFA becomes preferable to the ITM, which indicates that these techniques are complementary.

\subsection{Oscillating field}\label{sec:osc}

Finally, we consider a time-dependent laser pulse with a subcycle structure:
\begin{equation}
A_x (t) = \frac{E_0}{\omega} F(\omega t) \sin \omega t,\quad A_y = A_z = 0,\label{eq:lcfa_osc_field}
\end{equation}
where $F(\eta)$ is an envelope function. In particular, we choose a smooth profile which has an extended plateau:
\begin{equation}
F (\eta)=
\begin{cases}
\sin^2 \big [ \frac{1}{2} (\pi N - |\eta|) \big ] &\text{if}~~\pi (N-1) \leq |\eta| < \pi N,\\
1 &\text{if}~~|\eta| < \pi (N-1),\\
0 &\text{otherwise},
\end{cases} \label{eq:envelope}
\end{equation}
where $N$ is the number of cycles, so the pulse duration is $T = 2 \pi N/\omega$.

Since the vector potential is no longer monotonic, there are multiple turning points $t_*$ that contribute to $n_{\boldsymbol{p},s}^{\text{(LCFA)}}$. Moreover, each contribution relates to the same value of $|E(t_*)|$ once $F(t_*) = 1$, so the naive summation of $\mathrm{exp} (-\pi \lambda_{\boldsymbol{p}}[E(t_*)])$ leads to number densities which exceed unity for sufficiently large $N$. This fact obviously contradicts the Pauli exclusion principle. In order to avoid this obstacle, we suggest that the individual terms are summed according to the rule
\begin{equation}
n_{\boldsymbol{p},s}^{(i+1)} = n_{\boldsymbol{p},s}^{(i)} + [1 - n_{\boldsymbol{p},s}^{(i)}] \mathrm{e}^{-\pi \lambda_{\boldsymbol{p}}[E(t_*^{(i+1)})]},\label{eq:lcfa_osc_summation}
\end{equation}
where $i=0,1,...,K-1$, the positions of the turning points obey $t_*^{(1)}<t_*^{(2)}<...<t_*^{(K)}$, and $n_{\boldsymbol{p},s}^{\text{(0)}} = 0$. The prescription~(\ref{eq:lcfa_osc_summation}) is given by the classical probability theory. The LCFA result $n_{\boldsymbol{p},s}^{\text{(LCFA)}} = n_{\boldsymbol{p},s}^{(K)}$ is now always less than $1$ and tends to $1$ with increasing $N$.

\begin{figure*}[t]
\center{\includegraphics[height=0.35\linewidth]{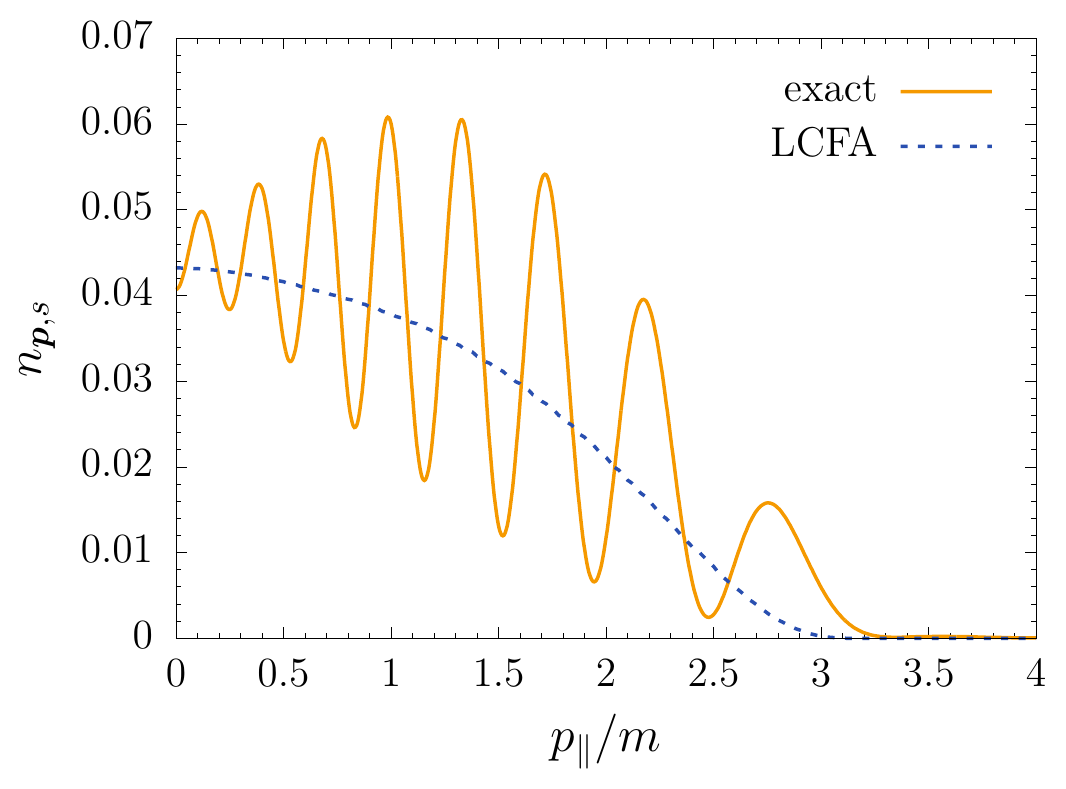}~~~~~\includegraphics[height=0.35\linewidth]{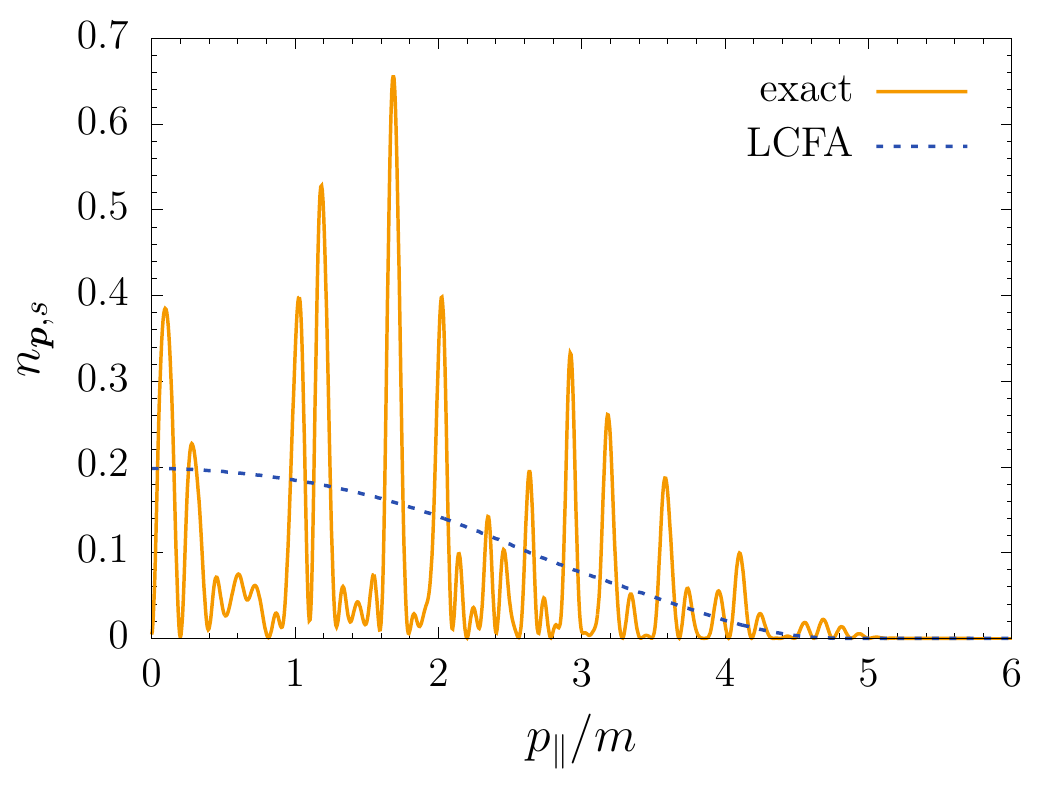}}
\caption{The momentum distribution of particles created by the external pulse~(\ref{eq:lcfa_osc_field}) for $N=1$ (left) and $N=3$ (right) ($E_0 = E_\text{c}$, $\omega = 0.2 m$, $p_\perp = 0$). The spectra are evaluated within the LCFA (dashed lines) and computed exactly (solid lines).}
\label{fig:osc_md}
\end{figure*}

This fact means that the LCFA does not describe the Rabi oscillations (the number density at given $\boldsymbol{p}$ oscillates as a function of the pulse duration), and we expect that the larger $N$ is, the less accurate predictions are made by the LCFA. In Fig.~\ref{fig:osc_md} we depict two examples of the momentum distributions using two different values of $N$.
One observes that the oscillating structure addressed in numerous studies (see, e.g., Refs.~\cite{akal_prd_2014, mocken_pra_2010, aleksandrov_prd_2017_1, fillion_pra_2012, abdukerim_plb_2013, dumlu_prl_2010, akkermans_prl_2012, kohlfuerst_prl_2014, hebenstreit_prl_2009}) is not reproduced by the LCFA as Eq.~(\ref{eq:lcfa_osc_summation}) does not take into account the interference among the different pair-production channels~\cite{comment_ITM_interference}. Moreover, the LCFA performs much worse for larger $N$ as the resonant peaks rise in the spectrum. Our calculations demonstrate that the LCFA can only provide a ``mean'' curve which can be considered as an adequate prediction in the case of short pulses. In addition, we note that the LCFA also fails to reproduce the interference effects in the photon spectra in the context of nonlinear Compton scattering~\cite{ilderton_arxiv_2018, harvey_pra_2015}.

\begin{figure}[t]
\center{\includegraphics[width=0.95\linewidth]{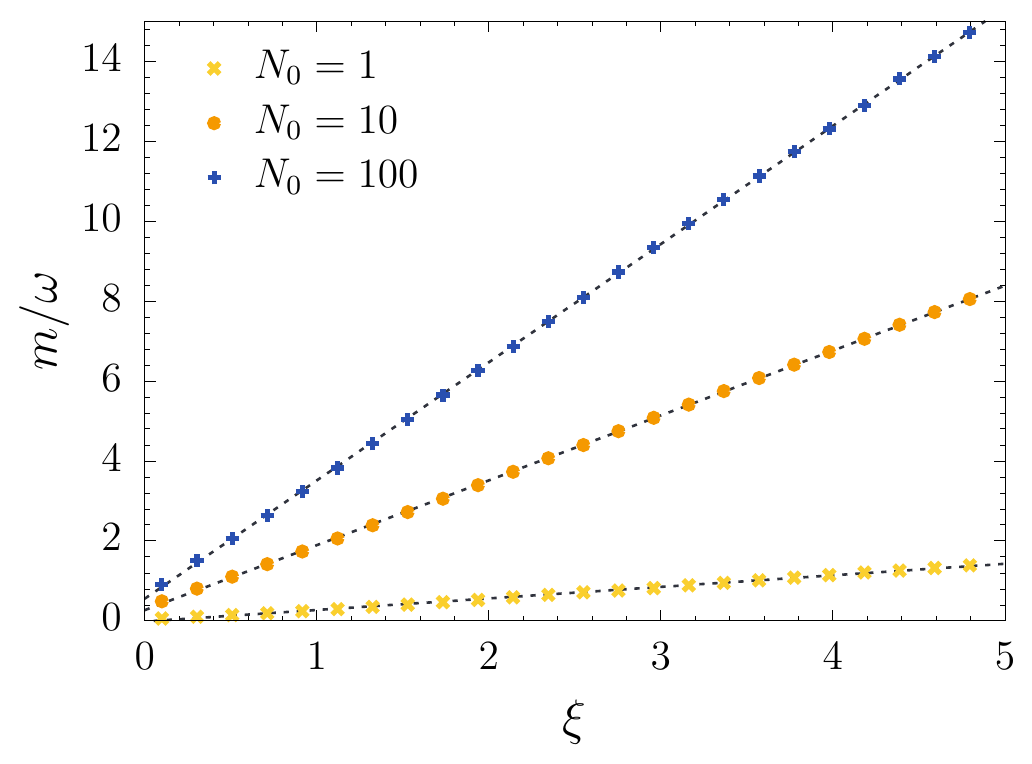}}
\caption{The ratio $m/\omega$ evaluated according to Eq.~(\ref{eq:lcfa_osc_omega_xi_N0}) as a function of $\xi$ for various $N_0$.}
\label{fig:omega_xi_N0}
\end{figure}

Let $\Omega_n$ be the Rabi frequency regarding the $n$th resonance [the $n$th peak has a height of $\sin^2 (\Omega_n T)$]. One has to require $\Omega_n T \ll \pi/2$ for all of the resonances in the momentum spectrum. This condition does not allow the resonances to form a pronounced peak structure. To formulate this requirement in terms of the laser field parameters $\xi$, $\omega$, and $N$, we set $\boldsymbol{p} = 0$ and calculate the Rabi frequencies for given $\xi$ and various resonance frequencies $\omega_n$. We introduce the characteristic number of cycles $N_n = \omega_n/(4\Omega_n)$ which yields the maximal number density of particles ($n_{\boldsymbol{p}=0,s} \approx 1$). In order to evaluate $\omega_n$ and $N_n$ as a function of $\xi$, one can turn to a quasiclassical treatment, as was done in Ref.~\cite{mocken_pra_2010} (see also Refs.~\cite{avetissian_pre_2002, kohlfuerst_prl_2014, kohlfuerst_prd_2018}). Let us introduce the approximate laser-dressed energy of the particle at rest ($\boldsymbol{p} = 0$), i.e. the effective mass:
\begin{equation}
q_0 = \frac{\omega}{2\pi} \int \limits_0^{2\pi/\omega} \sqrt{m^2 + e^2 A(t)^2} \, dt.\label{eq:lcfa_osc_energy}
\end{equation}
If one neglects the switching-on and -off parts of the laser pulse, where $F(\omega t) < 1$, one can recast Eq.~(\ref{eq:lcfa_osc_energy}) into
\begin{equation}
q_0 \approx \frac{m}{2\pi} \int \limits_0^{2\pi} \sqrt{1 + \xi^2 \sin^2 x} \, dx = \frac{m}{2\pi} \, \mathrm{E} (2\pi | -\xi^2), \label{eq:lcfa_osc_energy_E}
\end{equation}
where $\xi = |e|E_0/(m\omega)$ and $\mathrm{E} (z | k)$ is the incomplete elliptic integral of the second kind. The resonance condition now reads: $2q_0 = n \omega_n$. It turns out that for given $\xi$ the Rabi frequency of the $n$th resonance can be found via
\begin{equation}
\Omega_n = \frac{|e|E_0}{4\pi m} \Bigg | \int \limits_0^{2\pi} \frac{\cos x}{1+\xi^2 \sin^2 x} \, \mathrm{exp} \bigg [ \frac{2im}{\omega_n} \, \mathrm{E} (x | -\xi^2) \bigg ] \, dx \Bigg |.\label{eq:lcfa_osc_rabi_freq}
\end{equation}
The derivation of this equation is presented in Appendix~\ref{sec:appendix_1}.
In Fig.~\ref{fig:Nn_omega} we display the values of $N_n$ for various $n$ and $\xi$ plotting it versus $m/\omega_n$.
\begin{figure}[t]
\center{\includegraphics[width=0.95\linewidth]{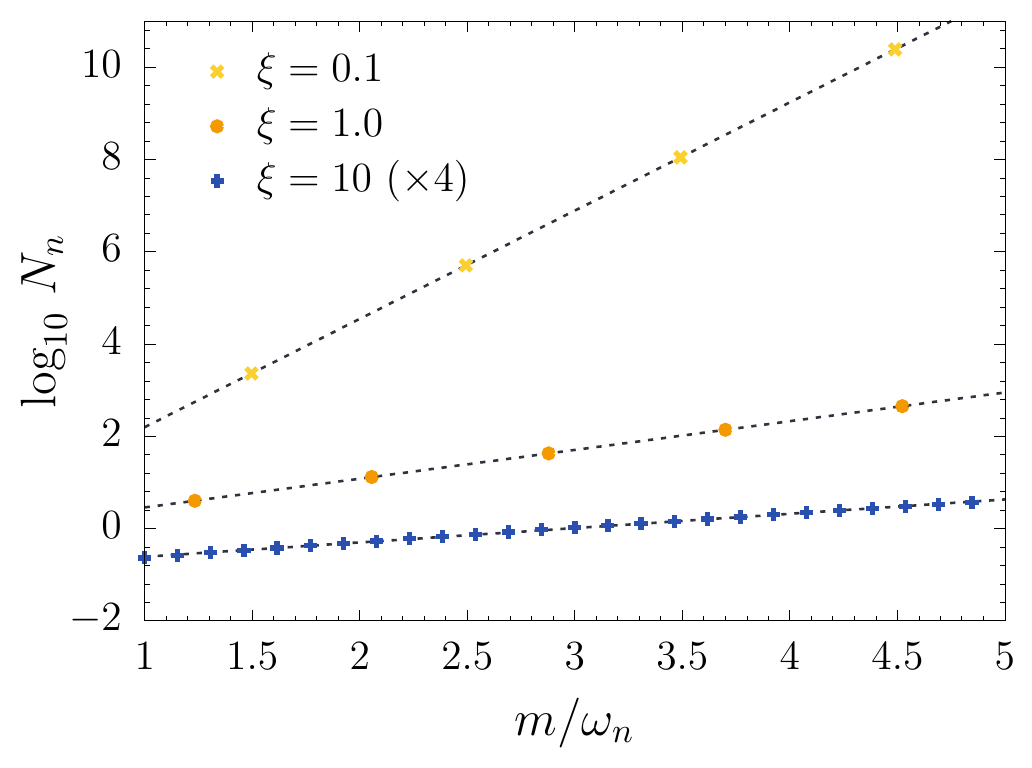}}
\caption{The number of cycles $N_n$ as a function of $\omega_n$ for various $\xi$. The results for $\xi = 10$ were multiplied by a factor of $4$.}
\label{fig:Nn_omega}
\end{figure}
Note that $n$ should always be odd due to the selection rule discussed, e.g., in Refs.~\cite{mocken_pra_2010, ruf_prl_2009, akal_prd_2014, aleksandrov_prd_2017_1, aleksandrov_prd_2018, kohlfuerst_prl_2014}. As was also shown in Ref.~\cite{mocken_pra_2010}, the results can be approximated according to
\begin{equation}
\ln N_n = a(\xi) + b(\xi)\, \frac{m}{\omega_n},\label{eq:lcfa_osc_Nn_omega_fit}
\end{equation}
which holds true for all $n$. Replacing $N_n$ and $\omega_n$ with continuous variables $N_0$ and $\omega$, respectively, one receives the characteristic number of cycles $N_0$ needed for the resonances to occur in the spectrum as a function of the field parameters $\xi$ and $\omega$. Let us then isolate $\omega$ as follows:
\begin{equation}
\frac{m}{\omega} = \frac{\ln N_0 - a(\xi)}{b(\xi)}.\label{eq:lcfa_osc_omega_xi_N0}
\end{equation}
In Fig.~\ref{fig:omega_xi_N0} we present the ratio $m/\omega$ as a function of $\xi$ for several different values of $N_0$.
One observes that the data can be fitted as
\begin{equation}
\frac{m}{\omega} = A(N_0) + B(N_0) \xi,\label{eq:lcfa_osc_omega_xi_N0_fit}
\end{equation}
which in turn leads to
\begin{equation}
\frac{\omega}{m} \, A(N_0) + \frac{E_0}{E_\text{c}} \, B(N_0) = 1.\label{eq:lcfa_osc_omega_xi_N0_line}
\end{equation}
This equation being considered at given $N_0$ yields a line in the $E_0$ -- $\omega$ plane. The line intersects the axes at $\omega/m = 1/A(N_0)$ and $E_0/E_\text{c} = 1/B(N_0)$, respectively. It turns out that the functions $A(N_0)$ and $B(N_0)$ depend linearly on $\log N_0$:
\begin{eqnarray}
A(N_0) &\approx& 0.123878 \, \ln N_0 - 0.015148, \label{eq:lcfa_osc_A}\\
B(N_0) &\approx& 0.579195 \, \ln N_0 + 0.288868. \label{eq:lcfa_osc_B}
\end{eqnarray}
These findings were confirmed by our direct numerical computations without using Eq.~(\ref{eq:lcfa_osc_rabi_freq}).

The results obtained should be interpreted as follows. Supposing that the external laser pulse has the parameters $E_0$ and $\omega$, one needs to find the value $N_0 = N_0 (E_0, \omega)$ which satisfies the condition (\ref{eq:lcfa_osc_omega_xi_N0_line}), i.e. the corresponding line should pass through the point $(E_0/E_\text{c}, \omega/m)$. To this end, one can use Eqs.~(\ref{eq:lcfa_osc_A}) and~(\ref{eq:lcfa_osc_B}). The resonant peaks in the momentum spectrum get close to their maxima as the number of cycles $N$ approaches $N_0 (E_0, \omega)$. Accordingly, the LCFA is expected to be adequate only if $N \ll N_0 (E_0, \omega)$. For instance, for the field parameters employed in Fig.~\ref{fig:osc_md}, one finds that $N_0 \approx 3.26$, which explains the appearance of pronounced resonant peaks already for $N = 3$.

Finally, we underline that even if all of the resonances are far from their maximal values, i.e. $N \ll N_0 (E_0, \omega)$, they can still considerably exceed the LCFA predictions. This means that the procedure described above provides only the necessary conditions for the field parameters. After fulfilling these requirements, one should directly compare the pair-production probabilities evaluated within the LCFA to the height of the possible resonant peaks in the spectrum, i.e. to $\sin^2 (\Omega_n T)$ for the corresponding values of $n$. This can also be done by means of Eq.~(\ref{eq:lcfa_osc_rabi_freq}), so the validity of the LCFA can be examined without performing the exact computations.

\section{Space-time-dependent fields}\label{sec:space-time}

In this section we will consider the case of a spatially inhomogeneous external field. We will first discuss how one can implement the LCFA for computing the momentum spectra of particles and then turn to benchmarking the LCFA predictions against the exact results.

\subsection{LCFA implementation}\label{sec:space_time_lcfa_gen}

First, we note that the presence of the spatial dependence substantially reduces the efficiency of the LCFA prescriptions formulated in the previous section. Although the particle momentum can be easily propagated in time in the case of uniform external fields, this task becomes much more complicated once some spatial inhomogeneities take place. Furthermore, one now needs not only to solve the equations of motion, but also to integrate over the possible values of the final position of the particle. Besides, the Pauli exclusion principle should also be taken into account, which makes the evaluation of the momentum spectra considerably difficult despite the approximate character of the computations. An alternative approach suggests that one calculates the pair-production probabilities replacing the external field with a spatially uniform background whose temporal dependence coincides with that of the original field configuration at a given position in space and sums then the results over the spatial region where the external field is present.

We will show now that the differential probabilities calculated according to this approach and integrated then over momentum provide the conventional LCFA formula for the total particle yield (see, e.g., Refs.~\cite{narozhny_bulanov, bulanov_prl}). We assume for simplicity that the external field points along the $x$ direction and has the form $E(t, x) = - \partial_t A(t, x)$ and $E (t, x) \geq 0$ for all $t$ and $x$. For a given value of $x$, one can employ the LCFA approach discussed in Sec.~\ref{sec:uniform}, i.e. approximate the particle number density as
\begin{equation}
n_{\boldsymbol{p},s}^{\text{(LCFA)}} (x) = \mathrm{e}^{-\pi \lambda_{\boldsymbol{p}}\{E[t_* (p_\parallel),x]\}}, \label{eq:lcfa_dens}
\end{equation}
where $t_* (p_\parallel)$ is the solution of the equation $p_\parallel - e[A(t_*, x) - A(t_\text{out}, x)] = 0$ and we assume that $p_\parallel \in [e \{ A(t_\text{in}, x) - A(t_\text{out}, x)\}, 0]$. Since the function $A(t,x)$ is a monotonic function of $t$, there is a one-to-one correspondence between $t_*$ and $p_\parallel$. Having evaluated the expression~(\ref{eq:lcfa_dens}) for given $x$, we integrate then over $x$:
\begin{equation}
\frac{d N^{\text{(LCFA)}}_{\boldsymbol{p},s}}{d^3 \boldsymbol{p}} = \frac{S}{(2\pi)^2} \int \limits_{-\infty}^{+\infty} \frac{dx}{2\pi} \, \mathrm{e}^{-\pi \lambda_{\boldsymbol{p}}\{E[t_* (p_\parallel),x]\}}, \label{eq:lcfa_x_sum}
\end{equation}
where $S$ is the $yz$ cross section of the system. To obtain the total number of pairs produced, we first integrate over $p_\parallel$. This integration can be performed in terms of $t_*$ having in mind that $dp_\parallel = |e| E(t_*, x) dt_*$. Omitting the star, we receive
\begin{equation}
\frac{d N^{\text{(LCFA)}}_{\boldsymbol{p}_\perp,s}}{d^2 \boldsymbol{p}_\perp} = \frac{S}{(2\pi)^2} \int \limits_{-\infty}^{+\infty} \frac{dx}{2\pi} \, \int \limits_{t_\text{in}}^{t_\text{out}} dt \, |e| E(t,x) \, \mathrm{e}^{-\pi \lambda_{\boldsymbol{p}}[E(t,x)]}. \label{eq:lcfa_x_sum_ppar}
\end{equation}
Finally, we integrate over $\boldsymbol{p}_\perp$ using the explicit form of $\lambda_{\boldsymbol{p}}$ [Eq.~(\ref{eq:T_schwinger})] and take into account the spin factor $2$:
\begin{equation}
N^{\text{(LCFA)}} = \frac{S}{4\pi^3} \int \limits_{-\infty}^{+\infty} dx \int \limits_{t_\text{in}}^{t_\text{out}} dt \, e^2 E^2 (t,x) \, \mathrm{e}^{-\pi m^2/|eE(t,x)|}. \label{eq:lcfa_x_full}
\end{equation}
This result exactly coincides with the prediction of the LCFA developed for calculating the total amount of pairs produced~\cite{narozhny_bulanov, bulanov_prl} (see also Ref.~\cite{gavrilov_prd_2017}). However, we will focus on the momentum distribution of particles which can be calculated by means of Eq.~(\ref{eq:lcfa_x_sum}). Note that due to the monotonicity of the vector potential, one can calculate the number density for given $x$ with the aid of Eq.~(\ref{eq:lcfa_dens}) instead of using the prescription~(\ref{eq:lcfa_osc_summation}), which takes into account the Pauli exclusion principle. It becomes now clear that in the case of an arbitrary temporal dependence of the external field, none of the expressions (\ref{eq:lcfa_x_sum}) and (\ref{eq:lcfa_x_full}) incorporates Pauli blocking. Moreover, the integral over $x$ in these formulas {\it independently} sums the contributions corresponding to different values of $x$, which could lead to additional overestimation of the pair-production probabilities. A proper inclusion of the Pauli principle can be performed only within the exact multidimensional QED treatment.

In what follows, we will compare the LCFA predictions with the exact spectra of particles. Since the uniform-field problem was already discussed in Sec.~\ref{sec:uniform}, we will focus on the role of the spatial inhomogeneities using the exact results for the purely time-dependent configurations instead of the approximate integrand in Eq.~(\ref{eq:lcfa_x_sum}). Supposing that one can carry out the precise calculations for arbitrary $E_x (t)$, we will discuss how accurate the LCFA can perform in the presence of a non-uniform field $E_x (t, x)$. We will examine several specific field configurations benchmarking the LCFA results against the exact spectra which are obtained with the aid of our nonperturbative numerical approach described in Ref.~\cite{aleksandrov_prd_2016} (it was also applied in Refs.~\cite{aleksandrov_prd_2017_2,aleksandrov_prd_2018}).

We assume that the external field has the form
\begin{equation}
E_x (t, x) = E_0 \, \mathcal{G} (t) \mathcal{F} (x), \label{eq:x_field_gen}
\end{equation}
where the temporal and spatial profiles will be specified below.

\subsection{Uniform static field inside a capacitor of finite size}\label{sec:space_time_capacitor}

\begin{figure}[t]
\center{\includegraphics[width=0.95\linewidth]{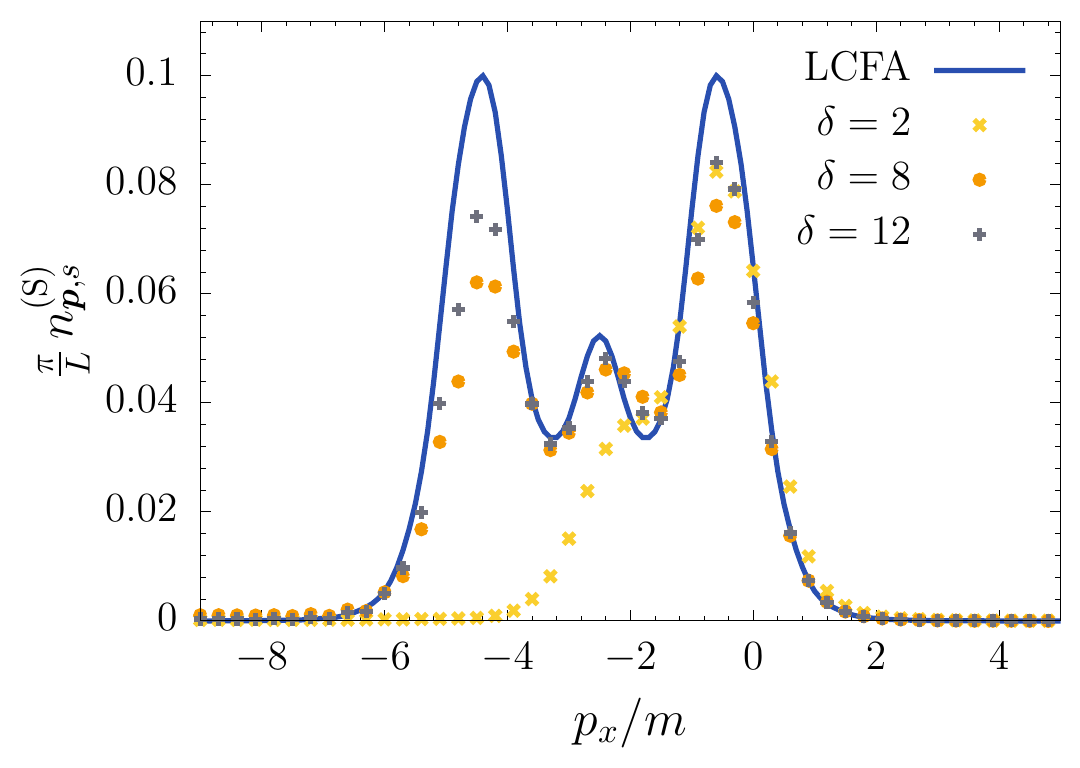}}
\caption{The momentum distribution of electrons created by the external field of the form~(\ref{eq:TL-const}) for $E_0 = E_\text{c}$, $T = 5 m^{-1}$, and various values of $\delta$ (points). The solid line represents the LCFA prediction.}
\label{fig:TL}
\end{figure}

First, we consider the case of a rectangular-like temporal and spatial profiles:
\begin{equation}
\mathcal{G} (t) = \theta (T/2 - |t|),\quad \mathcal{F} (x) = \theta (L - |x|). \label{eq:TL-const}
\end{equation}
For further convenience, we introduce the notations
\begin{equation}
\Pi = |e| E_0 \int \limits_{-\infty}^{+\infty} \mathcal{G} (t) dt, \quad L=\frac{\pi}{\Pi} \, \delta. \label{eq:TL_notations}
\end{equation}
Since the field configuration is now finite in the $x$ direction, the computations provide the following (finite) quantity:
\begin{equation}
n^\text{(S)}_{\boldsymbol{p},s} = \frac{(2\pi)^2}{S} \, \frac{d N_{\boldsymbol{p},s}}{d^3 \boldsymbol{p}}. \label{eq:density_notation_2D}
\end{equation}
We normalize the results multiplying them by a factor of $2\pi/(2L)$. After this renormalization, the summation over the $x$ coordinate leads exactly to the infinite-capacitor results discussed in the previous section which are to be compared with the exact values. As an example, we present the longitudinal momentum distributions for $E_0 = E_\text{c}$, $T = 5m^{-1}$, and various $\delta$ (see Fig.~\ref{fig:TL}). The transversal momentum equals zero, i.e. $\pi_\perp = m$.
The graph reveals indeed that the spectra found for the 2D field configuration recover the 1D result as $\delta \to \infty$. However, for small $\delta$ the effects of spatial finiteness become crucial, which means that the LCFA is well justified only for sufficiently large $\delta$. In Fig.~\ref{fig:TL} one observes that in the case $\delta = 2$, the spectrum support is strongly different from that obtained for large $\delta$. This can be understood if one notices that a classical particle in such a field configuration can escape from the region $x \in [-L,~L]$ before the field gets switched off. The left edge of the spectrum is formed by the particles produced at the very onset of the pulse, i.e. at $t = t_\text{in}$. If one requires the particle be still present inside the capacitor by the time instant $t = t_\text{out}$, it yields the condition $\delta \gg \delta_0$, where
\begin{equation}
\delta_0 = \frac{\Pi}{2\pi} \int \limits_{t_\text{in}}^{t_\text{out}} \frac{|e|E_0[\mathcal{A}(t_\text{in}) - \mathcal{A}(t)]}{\sqrt{\pi_{\perp}^2 + e^2 E_0^2 \big [ \mathcal{A} (t_\text{in}) - \mathcal{A}(t) \big ]^2}} \, dt
\label{eq:TL_condition}
\end{equation}
and $\mathcal{A}(t) = \int \limits^{t} \mathcal{G} (t') dt'$. In the case of a rectangular-like temporal profile, one obtains $\Pi = |e|E_0 T$ and
\begin{equation}
\delta_0 = \frac{\pi_{\perp} T}{\pi} \, \big [ \sqrt{1 + e^2 E_0^2 T^2/\pi_{\perp}^2} - 1 \big ].
\label{eq:TL_delta_0_explicit}
\end{equation}
For the field parameters from Fig.~\ref{fig:TL} and $\pi_{\perp} = m$, it yields $\delta_0 \approx 6.52$. Using Eq.~(\ref{eq:TL_delta_0_explicit}), one can now approximately identify the domain of the LCFA justification. The condition~(\ref{eq:TL_delta_0_explicit}) was derived within relativistic mechanics. The nonrelativistic regime appears once
\begin{equation}
\frac{|e|E_0 T}{m} \ll 1,
\label{eq:TL_nonrel}
\end{equation}
which leads to
\begin{equation}
\delta_0 \approx \frac{e^2 E_0^2 T^3}{2 \pi \pi_\perp}.
\label{eq:TL_nonrel_delta_0}
\end{equation}

The expressions (\ref{eq:TL_condition}) and ~(\ref{eq:TL_delta_0_explicit}) can be applied only in the case of slow spatial variations of the external potential and sufficiently large momentum of particles at the left edge of the spectrum. This requirement can be represented in the following form:
\begin{equation}
\Pi^3 \gg |e| E_0 \, \sqrt{m^2 + \Pi^2}.
\label{eq:TL_classic}
\end{equation}
In the nonrelativistic limit, it reads
\begin{equation}
\frac{e^2 E_0^2 T^3}{m} \gg 1.
\label{eq:TL_nonrel_classic}
\end{equation}
The derivation of the nonrelativistic form of the condition~(\ref{eq:TL_classic}) can be found, e.g., in Ref.~\cite{landau}, and its relativistic generalization leading to Eq.~(\ref{eq:TL_classic}) is quite straightforward.

We also note that in contrast to the results of Secs.~\ref{sec:uniform_rect} and \ref{sec:uniform_sauter}, the LCFA now overestimates the total number of particles. This can be explained using the fact that a locally constant treatment differently affects the particle yield depending on whether it is applied to the temporal dependence or spatial inhomogeneities. In the former case, the LCFA does not take into account ``dynamical production'' of particles due to fast variations of the external field (e.g., rapid switching on and off). However, in the latter case, the LCFA treats the particle as if it were interacting with a uniform and infinite background and thus prevents the particle from escaping, which leads to the overestimation in Fig.~\ref{fig:TL}. Note that these two patterns are clearly seen in our results since we treat here the temporal dependence exactly and therefore disentangle the two effects.

In what follows, we will discuss a Sauter-like temporal profile.

\subsection{Smooth temporal profile}\label{sec:t_sauter}

\begin{figure}[]
\center{\includegraphics[width=0.95\linewidth]{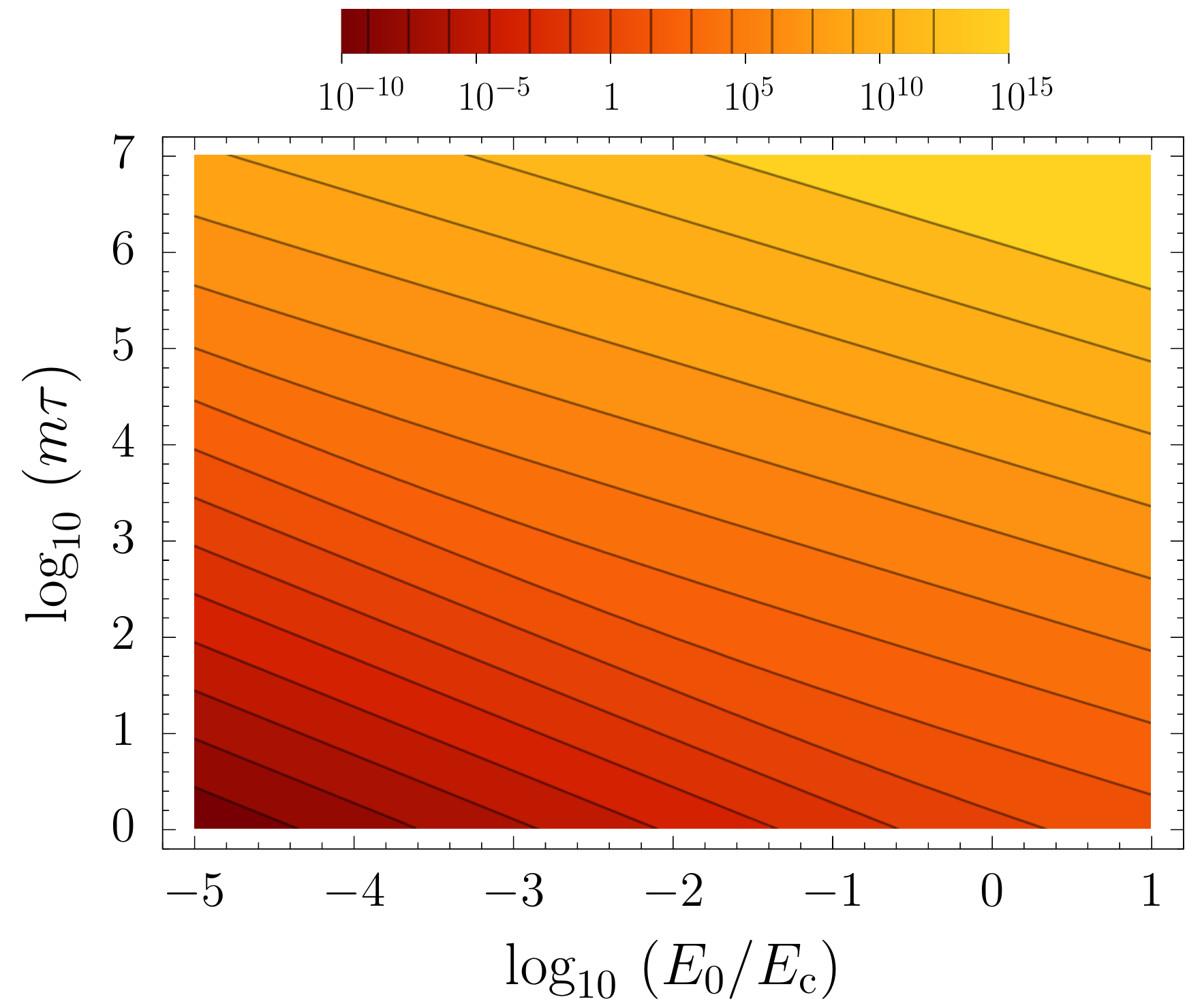}}
\caption{The dimensionless parameter $\delta_0$ for the field configuration~(\ref{eq:t_sauter}) as a function of $E_0$ and $\tau$.}
\label{fig:t_sauter_delta}
\end{figure}

The external field configuration now has the form
\begin{equation}
\mathcal{G} (t) = \cosh^{-2} (t/\tau),\quad \mathcal{F} (x) = \theta (L - |x|). \label{eq:t_sauter}
\end{equation}
This leads to $\Pi = 2|e|E_0 \tau$. Although in this case $t_\text{in/out} \to \mp \infty$, the integral in Eq.~(\ref{eq:TL_condition}) should not be computed along the whole axis. Indeed, the field exerts a non-negligible force on the particle only within the interval $|t| \lesssim \tau$. Accordingly, in Eq.~(\ref{eq:TL_condition}) one should replace $t_\text{in/out}$ with $\mp \tau$, respectively. In Fig.~\ref{fig:t_sauter_delta} we display the dependence of $\delta_0$ on $E_0$ and $\tau$.
Although the corresponding integral can be calculated exactly, we present only the asymptotic behavior due to the cumbersomeness of the full expression. For $\xi \ll 1$, one obtains
\begin{equation}
\delta_0 \approx \frac{2 e^2 E_0^2 \tau^3}{\pi \pi_{\perp}} \, \tanh 1, \quad \frac{8 e^2 E_0^2 \tau^3}{m} \gg 1. \label{eq:t_sauter_asym_small}
\end{equation}
The inequality shown in Eq.~(\ref{eq:t_sauter_asym_small}) is derived from Eq.~(\ref{eq:TL_classic}). For $\xi \gg 1$, we receive
\begin{equation}
\delta_0 \approx \frac{|e|E_0\tau^2}{\pi},\quad 4 |e| E_0 \tau^2 \gg 1. \label{eq:t_sauter_asym_large}
\end{equation}
For large $\pi_\perp / m$ one should replace $\xi$ with $|e|E_0\tau / \pi_\perp$ in the conditions $\xi \gg 1$ and $\xi \ll 1$. Note that the temporal-profile width $\tau$ in the strong-coupling regime $\xi \gg 1$ should obey $\tau \gg 1/(4\xi m)$ which is always satisfied by realistic pulse durations. Moreover, if $\xi \gtrsim 1$ but it does not fulfill $\xi \gg 1$, Eq.~(\ref{eq:TL_classic}) requires $|e|E_0 \ll m^2$ which is also completely realistic from the experimental viewpoint. It is a crucial point since it indicates that the semiclassical analysis of the particle trajectories is always justified once the nonperturbative pair-production process is considered. The expressions (\ref{eq:t_sauter_asym_small}) and (\ref{eq:t_sauter_asym_large}) demonstrate again that the field parameters must obey nontrivial relations to make the LCFA results valid.

\subsection{Smooth temporal and spatial profiles}\label{sec:tx_sauter}

\begin{figure}[t]
\center{\includegraphics[width=0.95\linewidth]{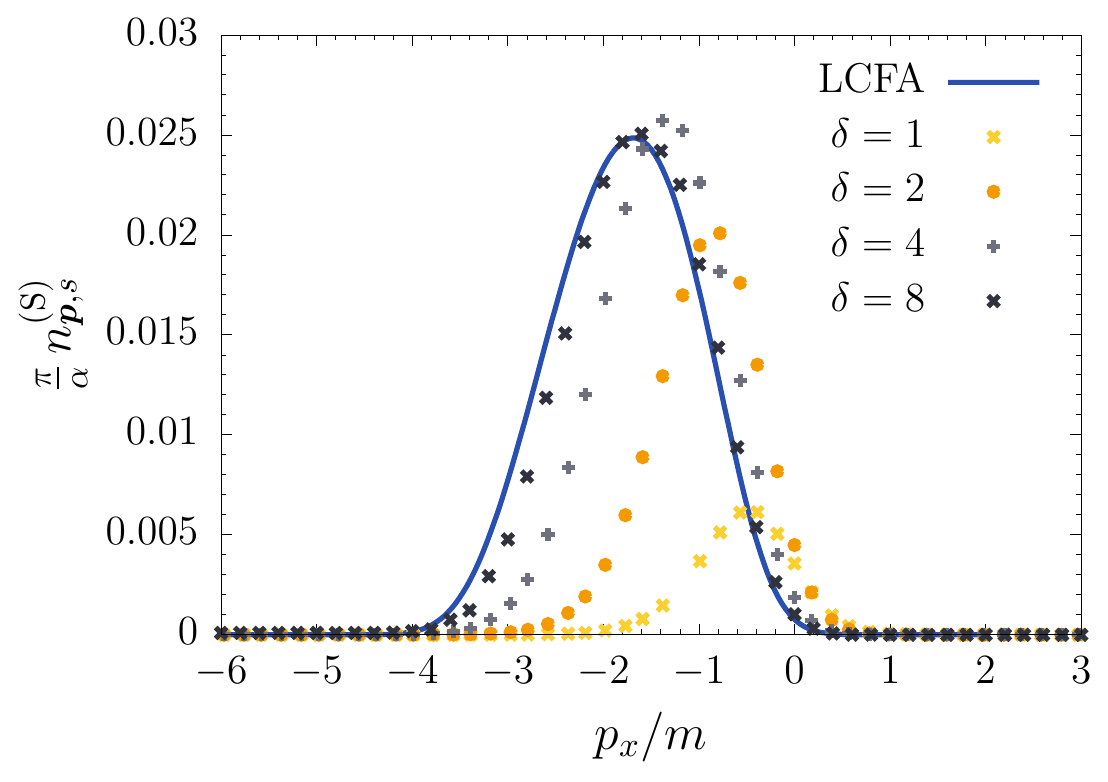}}
\caption{The momentum distribution of electrons produced by the external field with a smooth spatiotemporal profile~(\ref{eq:tx_sauter}) for $E_0 = E_\text{c}$, $\tau = 2 m^{-1}$, and various values of $\delta$ (points). The solid line represents the spectrum obtained within the LCFA.}
\label{fig:tx_example}
\end{figure}

\begin{figure}[t]
\center{\includegraphics[width=0.95\linewidth]{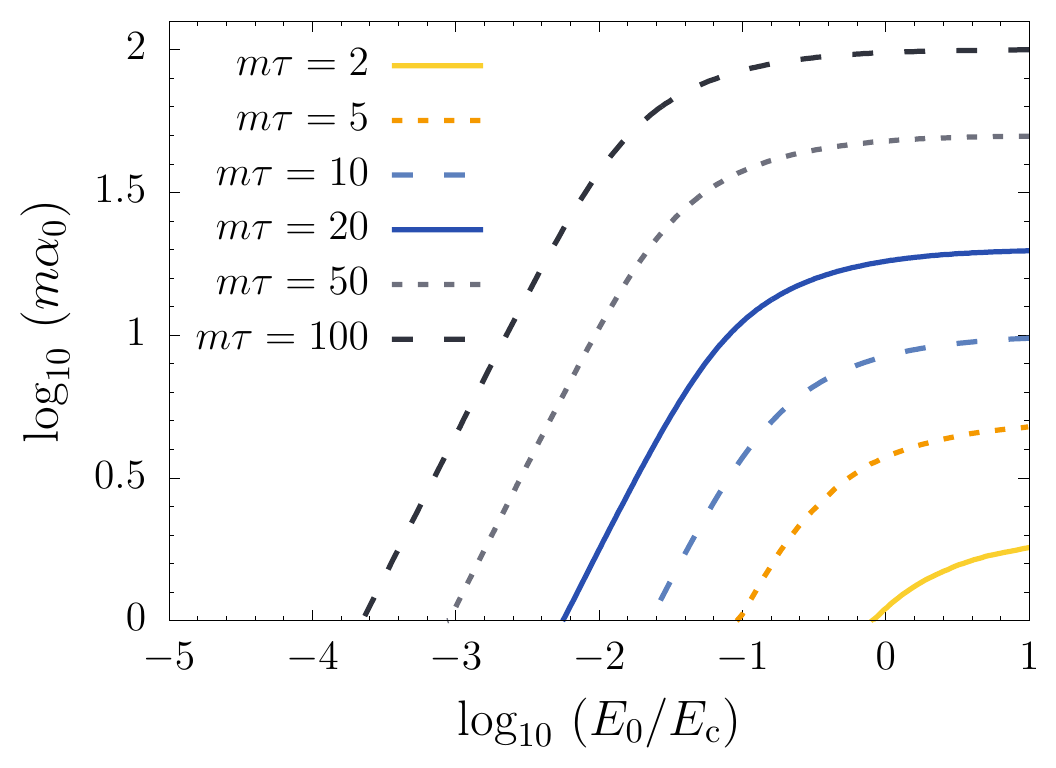}}
\caption{The dependence of the parameter $\alpha_0$ on the peak field strength $E_0$ and the pulse duration $\tau$.}
\label{fig:tx_alpha_0}
\end{figure}

Finally, we consider a smooth spatial profile:
\begin{equation}
\mathcal{G} (t) = \cosh^{-2} (t/\tau),\quad \mathcal{F} (x) = \cosh^{-2} (x/\alpha). \label{eq:tx_sauter}
\end{equation}
In the case of this configuration, the summation over the spatial coordinate within the LCFA becomes more complicated. One should now integrate the exact expression~(\ref{eq:lcfa_sauter_exact}) varying the parameter $E_0$ according to $E_0 (x) = E_0 \mathcal{F} (x)$, where $-\infty < x < +\infty$. A simple substitution of the integration variable $\tilde{x} = x/\alpha$ demonstrates that the result of this integration being divided by $\alpha$ is independent of $\alpha$. Accordingly, we divide the number density by $\alpha / \pi$, so that one can compare the exact results with the LCFA spectrum. We introduce the notation
\begin{equation}
\alpha = \frac{\pi}{\Pi} \, \delta,~~\text{where}~~\Pi = 2|e|E_0 \tau, \label{eq:tx_alpha_delta}
\end{equation}
and depict the spectra found within the LCFA and computed exactly for $E_0 = E_\text{c}$, $\tau = 2 m^{-1}$, and various values of $\delta$ (see Fig.~\ref{fig:tx_example}).
The qualitative behavior of the exact spectra for different $\delta$ is similar to what was reported in Ref.~\cite{hebenstreit_prl_2011}: the spectra shift along the $p_x$ axis, and the particle yield vanishes as $\delta \to 0$. We observe that the LCFA performs accurately only for sufficiently large $\delta$. To describe this behavior, one can perform again the analysis of the classical trajectories of the relativistic particle in the external field. Setting $\pi_\perp = m$ and solving the classical equations of motion numerically for various $E_0$, $\tau$, and $\alpha$, we calculate the characteristic spatial width $\alpha_0$ corresponding to the trajectories starting at $x = \alpha$ (with zero velocity) and ending at $x = -\alpha$ as the temporal variable changes from $-\tau$ to $\tau$. In Fig.~\ref{fig:tx_alpha_0} this quantity is presented as a function of $E_0$ for several different values of $\tau$.
For the values employed in Fig.~\ref{fig:tx_example}, it gives $\alpha_0 \approx 1.12 m^{-1}$, i.e. $\delta_0 \approx 1.42$. Fitting the data obtained, one can identify the following scaling with respect to $E_0$ and $\tau$ in the limit $\xi \ll 1$:
\begin{equation}
\alpha_0 \sim \frac{|e|E_0\tau^2}{m}, \qquad \delta_0 \sim \frac{e^2 E_0^2 \tau^3}{m}.\label{eq:tx_asym_small}
\end{equation}
For $\xi \gg 1$, one obtains
\begin{equation}
\alpha_0 \sim \tau, \qquad \delta_0 \sim |e|E_0 \tau^2.\label{eq:tx_asym_large}
\end{equation}
We observe now that the scaling of the parameter $\delta_0$ exhibits a universal behavior [compare Eqs.~(\ref{eq:tx_asym_small}) and~(\ref{eq:tx_asym_large}) with Eqs.~(\ref{eq:t_sauter_asym_small}) and~(\ref{eq:t_sauter_asym_large})]. It means that the shape of the spatial profile of the external field does not play here a major role.

Finally, we stress that the requirement that the classical particle be confined in the vicinity of the field maximum (in this case it means $-\alpha \lesssim x \lesssim \alpha$) is not equivalent to the condition $l_\text{c} = m / (|e|E_0) \ll \alpha$, where $l_\text{c}$ is the characteristic pair-formation length. The latter does not take into account the particle dynamics in the presence of the external field. Nevertheless, replacing the electron mass in the expression for $l_\text{c}$ with the relativistic energy of the particle with momentum $\Pi$, one receives the condition~(\ref{eq:tx_asym_large}) in the limit $\xi \gg 1$. On the other hand, the more extensive analysis of the particle trajectories conducted in this section represents a more general tool for justifying the LCFA.

\section{Discussion}\label{sec:discussion}

In the present study, we analyzed a number of simple configurations of the external electric field in order to benchmark the locally-constant field approximation against the exact methods and deduce the requirements that should be fulfilled if one aims at utilizing the LCFA in one's calculations. In particular, we focused on the momentum distributions of particles produced. The first part of the study was devoted to the case of a spatially-uniform electric field. It was shown that the criteria of the LCFA applicability turn out to be rather nontrivial even if very simple temporal profiles of the external field are considered. For instance, in the case of a Sauter pulse, the momentum spectrum can be accurately described by the LCFA only when $|eE_0|^{3/2} \tau \gg m^2$. This condition is much stronger than $\xi = |e| E_0 \tau/m \gg 1$, so the LCFA can be justified only in the deeply nonperturbative regime. Next we turned to the analysis of an oscillating field profile giving rise to multiple turning points of the classical-particle motion. Since the LCFA does not capture the oscillating structure of the momentum spectra, it can be invoked only for studying short laser pulses, i.e. pulses containing sufficiently small number of cycles. In order to clarify this issue, we focused on the resonant Rabi oscillation and evaluated the corresponding Rabi frequency as a function of the pulse amplitude and frequency. It was demonstrated that performing a quite simple analysis of the $n$-photon resonances, one can find out whether the LCFA should yield reliable predictions. In the second part of the present investigation, we examined several non-uniform external backgrounds. It was shown that the LCFA may indeed perform well as long as the corresponding classical trajectories are localized within the spatial region where the external field is close to its maximum. This provides a generic approach which can be used in the preliminary examination of the external field configuration before the LCFA is employed. Besides, it was found that the validity of the LCFA is not sensitive to the details of the field spatial profile. What could be even more important is the fact that the estimates extracted from the properties of the classical trajectories should be accurate once one is interested in the strong-coupling regime $\xi \gtrsim 1$.

Although the present investigation involved the simplest field configurations, the corresponding findings can already provide valuable insights into the LCFA justification in the case of more realistic scenarios. First, the results of Sec.~\ref{sec:uniform} indicate that the temporal dependence of the external background can hardly be taken into account within the LCFA as the real laser setups may well contain too many carrier cycles while this approximation does not take into account the multiphoton signatures in the momentum spectra. However, the exact treatment of the temporal dependence of the external field and further summation over the spatial coordinates could still efficiently provide quite accurate results. To judge whether this summation leads to adequate predictions, one can examine the particle dynamics similarly to what was discussed in Sec.~\ref{sec:space-time}. The aforementioned criterion formulated in terms of the classical trajectories can be applied in the case of an arbitrary field configuration, provided $\xi \gtrsim 1$. In particular, this treatment is expected to further illuminate how the magnetic field component affects the validity of the LCFA. Moreover, the justification of more sophisticated modifications of the LCFA approach (see, e.g., Ref.~\cite{kohlfuerst_epjp_2018}) can also be addressed by means of similar considerations. To carefully explore these ideas, one has to conduct the calculations for more complex external backgrounds, which is an important task for future studies.

Finally, we point out that the LCFA can also be employed for the approximate evaluation of the total number of pairs produced. Since in this case one does not need to follow the momentum of the particle once it is created by the external field, it is easier to suggest the corresponding approximation for this integral quantity [for instance, see Eq.~(\ref{eq:lcfa_x_full})]. Furthermore, this simplification imposes weaker restrictions on the field parameters. For instance, the evaluation of the total particle yield in the case of a Sauter pulse can be accurately performed even if the condition~(\ref{eq:lcfa_sauter_condition}) is not satisfied~\cite{gavrilov_prd_2017}. Benchmarking this kind of the LCFA approach is beyond the scope of the present investigation.

\section*{Acknowledgments}

This  investigation  was  supported  by  Russian  Foundation for   Basic   Research   (RFBR)   and   Deutsche Forschungsgemeinschaft (DFG) (Grants No. 17-52-12049 and No. PL 254/10-1) and by Saint Petersburg State University (SPbSU) and DFG (Grants No. 11.65.41.2017 and No. STO 346/5-1). I.~A.~A. also acknowledges the support from the FAIR-Russia Research Center and from the Foundation for the advancement of theoretical physics and mathematics ``BASIS''.

\appendix*
\section{Rabi frequency}\label{sec:appendix_1}

In the case of a monochromatic spatially-uniform external field, the frequencies of the $n$-photon resonances for $\boldsymbol{p} = 0$ can be found according to $\omega_n = 2q_0 / n$, where the quasienergy $q_0$ is given by Eq.~(\ref{eq:lcfa_osc_energy}). The number density of particles produced then oscillates as a function of the pulse duration: $n_{\boldsymbol{p}=0, s} \approx \sin^2 (\Omega_n T)$. The closed-form expression~(\ref{eq:lcfa_osc_rabi_freq}) for the corresponding Rabi frequency $\Omega_n$ can be derived by the quasiclassical consideration which is valid for $\omega \ll m$ and $E_0 \ll E_\text{c}$~\cite{mocken_pra_2010}.

Let us introduce the following function:
\begin{equation}
d(t) \equiv \mathcal{P} (t) \mathrm{e}^{2iS_0 (t)} \mathrm{e}^{-2iq_0 t},\label{eq:app_d_def}
\end{equation}
where
\begin{equation}
\mathcal{P} (t) = - \frac{ieE(t)m}{2 \varepsilon^2 (t)}, \quad \varepsilon (t) = \sqrt{m^2 + e^2 A^2 (t)},\label{eq:app_Pcal}
\end{equation}
and the action $S_0$ has the form
\begin{equation}
S_0 (t) = \int \limits^{t} \varepsilon (t') dt'.\label{eq:app_action}
\end{equation}
We set $\omega = \omega_n$. The function $d(t)$ is periodic, and it turns out that the $n$th Fourier coefficient provides the corresponding Rabi frequency~\cite{mocken_pra_2010}:
\begin{equation}
d(t) = \sum_k c_k \mathrm{e}^{-in\omega_n t}, \quad \Omega_n = |c_n|.\label{eq:app_Fourier}
\end{equation}
It follows that
\begin{equation}
c_n = -\frac{im\omega_n}{4\pi} \int \limits_0^{2\pi / \omega_n} \frac{eE(t)}{\varepsilon^2 (t)} \, \mathrm{e}^{2iS_0 (t)} dt.\label{eq:app_cn}
\end{equation}
Using the explicit form of the vector potential~(\ref{eq:lcfa_osc_field}) and neglecting the switching-on and -off parts of $F(\omega t)$, one obtains
\begin{equation}
S_0 (t) = \frac{m}{\omega_n} \, \mathrm{E} (\omega_n t | -\xi^2) + \text{\{real constant\}}.\label{eq:app_action_E}
\end{equation}
With the aid of Eqs.~(\ref{eq:app_cn}) and (\ref{eq:app_action_E}), one receives Eq.~(\ref{eq:lcfa_osc_rabi_freq}) for $\Omega_n = |c_n|$.


\end{document}